\documentclass[fleqn,usenatbib]{mnras}

\usepackage{newtxtext,newtxmath}

\usepackage[T1]{fontenc}

\DeclareRobustCommand{\VAN}[3]{#2}
\let\VANthebibliography\thebibliography
\def\thebibliography{\DeclareRobustCommand{\VAN}[3]{##3}\VANthebibliography}

\usepackage{graphicx}	
\usepackage{amsmath}	
\usepackage{academicons} 
\usepackage{xcolor}
\usepackage{url}
\usepackage[utf8]{inputenc}
\usepackage{hyperref}
\usepackage{orcidlink}
\usepackage{multicol}
\usepackage{pdflscape}

\newcommand{\kms}{\mbox{km s$^{-1}~$}} 
\newcommand{\kmse}{\mbox{km s$^{-1}$}}

\newcommand{\msun}{M$_{\odot}~$} 
\newcommand{\msune}{M$_{\odot}$}

\newcommand{\dgr}{$^{\circ}~$}

\newcommand{\teff}{$\rm T_{eff}$~}
\newcommand{\teffe}{$\rm T_{eff}$}
\newcommand{\logg}{$\log{g}$~}
\newcommand{\logge}{$\log{g}$}

\title[SMC Gradients]{Revealing the Chemical Structure of the Magellanic Clouds with APOGEE. III. Abundance Gradients of the Small Magellanic Cloud}

\author[Povick et al.]{
Joshua T. Povick\orcidlink{0000-0002-6553-7082}$^{1}$\thanks{E-mail: joshua.povick@montana.edu},
David L. Nidever\orcidlink{0000-0002-1793-3689}$^{1}$,
Pol Massana\orcidlink{0000-0002-8093-7471}$^{1}$,
Steven R. Majewski\orcidlink{0000-0003-2025-3147}$^{2}$,
Yuxi(Lucy) Lu\orcidlink{0000-0003-4769-3273}$^{3,4}$,
\newauthor
 Maria-Rosa L. Cioni\orcidlink{0000-0002-6797-696X}$^{5}$,
Doug Geisler$^{6,7,8}$,
Szabolcs~M{\'e}sz{\'a}ros\orcidlink{0000-0001-8237-5209}$^{9,10}$,
Christian Nitschelm\orcidlink{0000-0003-4752-4365}$^{11}$,
\newauthor
 Andr\'{e}s Almeida\orcidlink{0009-0000-0733-2479}$^{2}$,
Richard R. Lane\orcidlink{0000-0003-1805-0316}$^{12}$,
and Pen\'{e}lope Longa-Pe\~{n}a$^{11}$\\
$^{1}$Department of Physics, Montana State University, P.O. Box 173840, Bozeman, MT 59717-3840\\
$^{2}$Department of Astronomy, University of Virginia, Charlottesville, VA 22904-4325, USA \\
$^{3}$American Museum of Natural History, Central Park West, Manhattan, NY, USA \\
$^{4}$Department of Astronomy, Columbia University, 550 West 120th Street, New York, NY, USA \\
$^{5}$Leibniz-Institut f\"{u}r Astrophysik Potsdam, An der Sternwarte 16, D-14482 Potsdam, Germany \\
$^{6}$Departmento de Astronom\'{i}a, Universidad de Concepci\'{o}n, Casilla 160-C Concepci\'{o}n, Chile \\
$^{7}$Instituto Multidisciplinario de Investigaci\'{o}n y Postgrado, Universidad de La Serena, Ra\'{u}l Bitr\'{a}n 1305, La Serena, Chile\\
$^{8}$Departamento de Astronom\'{i}a, Facultad de Ciencias, Universidad de La Serena. Av.
Juan Cisternas 1200, La Serena, Chile \\
$^{9}$ELTE E\"{o}tv\"{o}s Lor\'and University, Gothard Astrophysical Observatory, 9700 Szombathely, Szent Imre H. st. 112, Hungary \\
$^{10}$MTA-ELTE Lend\"{u}let ``Momentum'' Milky Way Research Group, Hungary \\
$^{11}$Centro de Astronom{\'i}a (CITEVA), Universidad de Antofagasta, Avenida Angamos 601, Antofagasta 1270300, Chile \\
$^{12}$Centro de Investigaci\'{o}n en Astronom\'{i}a, Universidad Bernardo O'Higgins, Avenida Viel 1497, Santiago, Chile \\
}

\date{Accepted XXX. Received YYY; in original form ZZZ}

\pubyear{2023}

\begin{document}
\label{firstpage}
\pagerange{\pageref{firstpage}--\pageref{lastpage}}
\maketitle

\begin{abstract}

    We determine radial- and age-abundance gradients of the Small Magellanic Cloud (SMC) using spectra of 2,062 red giant branch (RGB) field stars observed by SDSS-IV / APOGEE-2S. With coverage out to $\sim$9 kpc in the SMC, these data taken with the high resolution ($R \sim 22,500$) APOGEE $H$-band spectrograph afford the opportunity to measure extensive radial gradients for as many as 24 abundance ratios. The SMC is found to have an overall metallicity gradient of $-$0.0546 $\pm$ 0.0043 dex/kpc. Ages are calculated for every star to explore the evolution of the different abundance gradients. As a function of age, many of the gradients show a feature 3.66--5.58 Gyr ago, which is especially prominent in the [X/H] gradients. Initially many gradients flatten until about $\sim$5.58 Gyr ago, but then steepen in more recent times. We previously detected similar evolutionary patterns in the Large Magellanic Cloud (LMC) which are attributed to a recent interaction between the LMC and SMC. It is inferred that the feature in the SMC gradients was caused by the same interaction. The age-[X/Fe] trends, which track average [X/Fe] over time, are flat, demonstrating a slow enrichment history for the SMC. When comparing the SMC gradients to the LMC and MW, normalized to disk scale length ($R_\text{d}$), the [X/Fe] and [X/Mg] gradients are similar, but there is a dichotomy between the dwarfs and the Milky Way (MW) for the [X/H] gradients. The median MW [X/H] gradient around $-$0.125 dex/$R_\text{d}$ whilst the Clouds have gradients of about $-$0.075 dex/$R_\text{d}$.

\end{abstract}




\section{Introduction}
\label{sec:smc_intro}

The Small and Large Magellanic Clouds (MCs) are a conjoined pair of massive dwarf satellite galaxies currently approaching our Milky Way (MW).
The Small Magellanic Cloud (SMC) has a stellar mass of $\sim$3--3.5 $\times$ 10$^8$ \msun \citep{skibba2012spatial} and dark matter mass of $M_{\mathrm{DM}}$($R$ $<$ 4 kpc) = 1--1.5 $\times 10^9$ \msun \citep{diTeodoro2019}. The galaxy is also well within the virial radius of the MW at a distance of 62.44 kpc \citep{graczyk2020distance}, and, therefore, relatively luminous,
with an integrated apparent $V$ magnitude of 2.2 $\pm$ 0.2 \citep{mcconnachie2012observed}. Combined, these factors make the SMC a prime candidate for resolved stellar population analyses with high-resolution spectroscopy,
because it presents a large number of 
accessible red giant branch (RGB) stars brighter than $V$ $\sim$ 18 \citep[e.g.,][]{Harris2006,Dobbie2014,nidever20lazy}.  

The SMC is a dwarf irregular galaxy (dIrr) with a complex geometry.
The entire shape of the galaxy has been heavily affected by tidal interactions with the Large Magellanic Cloud (LMC), and continues to be a challenge to model. It has been shown that the SMC is very extended along the line-of-sight on the eastern side (towards the LMC), more than the extent it shows face-on and as much as 20 kpc based on observations of standard candles \citep{Scowcroft2016,Nidever2013smc}.
Previously seen and described as a ``distance modality'' \citep{Nidever2013smc}, this deepening of the eastern SMC may represent tidal material pulled out of the SMC (e.g., Almeida et al. 2023, submitted).

One of the most striking characteristics of the MCs as a whole is the trail of HI gas coming off the MCs, known as the ``Magellanic Stream'' \citep[MS; e.g.,][]{Mathewson1974,Putman2003,Bruens2005,Nidever2008}, which was discovered to extend up to 200\dgr across the sky \citep{Nidever2010}. The Stream is thought to be made mostly of stripped SMC gas \citep{Besla2012,Lucchini2020} in a scenario where the MCs are in their first infall to the MW \citep{Besla2007,Kallivayalil2013}.

Interacting galaxies are 
typically characterized by tidally or collisionally incited burts of
star formation 
\citep[e.g.,][]{privon2017widespread}. The MCs are known to have had a close interaction in the recent past based on their observed kinematics \citep{Zivick2018, nidever20lazy,niederhofer2021kinematics,schmidt2022kinematics}. The Magellanic Bridge \citep[MB;][]{Hindman1963}, a span
of gas and young stars \citep[$\sim$50 Myr old;][]{Demers1991,Muller2007} between the two galaxies (stretching $\sim$10\dgr from the SMC to the LMC), has long been thought to be the result of their most recent interaction, only $\sim$200 Myr ago  \citep[e.g.,][]{Gardiner1994,Muller2007,Besla2012,Diaz2012,schmidt2020bridge}. Additionally, the MCs are expected to have been interacting prior to that event, based on the modelling of their tidal debris \citep{Cullinane2022} and star formation histories \citep{rubele2018sfh,mazzi2021sfh,Massana2022}. Therefore, it is logical to expect that the chemical evolution 
of the MCs has been significantly influenced by periodic 
interactions and bursts of star formation. Naturally, the chemical abundance gradients in either galaxy should be expected to respond 
as a consequence.


Among the attempts to explore the recent close interaction of the MCs, 
\citet{nidever20lazy} used the \texttt{flexCE} code \citep{andrews2017inflow} to model the chemical evolution of the LMC, and, by this means, show evidence for a spike in star formation in the LMC about 2 Gyr ago.
This was accomplished by allowing the the star formation efficiency (SFE) to vary as a function of time and superimposing a Gaussian peak on top of the SFE to act as the burst.
Using similar modeling with \texttt{flexCE} and a time-evolving SFE, \cite{hasselquist2021satellites} was able to confirm the timeframe of the star formation burst in the LMC.
That work also modelled the chemical evolution of the SMC and found that the burst was weaker and happened $\sim$3--4 Gyr earlier than in the LMC. 
Based on this apparent delay between the star formation bursts in the MCs due to their close interaction, it is expected that changes in the chemistry of the SMC should predate analogous changes in the LMC. 



Interacting dwarf galaxy systems are important tools for
understanding how the MW and other larger galaxies formed. The prevailing ``hierarchical merging'' galaxy formation paradigm 
\citep{white1978merge,searle78} is that
galaxies 
grow through a steady accumulation of merging subsystems on a growing spectrum of mass scales.
Smaller, dwarf galaxies were the first type of galaxy to form, and, then 
these merged to form galaxies in the large range of sizes and masses that we see today. However, our knowledge of the details of these earlier processes is still incomplete. While even today the Universe is 
full of dwarf galaxies still engaged in the process of hierarchical merging, detailed study of this process is typically challenged by the fact that most prime examples are too faint and distant to be in the realm of resolved stellar population studies.  For this reason, the MCs present a particularly unique opportunity to dissect --- with star-by-star, high resolution spectroscopic analysis ---
a key step in the process of galaxy assemblage.

Previously, most spectroscopic studies of SMC stars could only explore its overall metallicity gradient \citep[e.g.][]{parisi2010smcgrad,dobbie2014smcgrad,parisi2016smcgrad,choudhury2020smcgrad,parisi2022cat,debortoli2022smcgrad,li2023photometric} and not other abundances. 
This is because multi-element chemical abundance 
studies require high $S/N$, high-resolution spectra to derive precise elemental abundances, whereas the mean metallicity of stars is more readily measured with lower resolution ($R\sim1000$), lower $S/N$ data.  
However, the Apache Point Observatory Galactic Evolution Experiment \citep[APOGEE,][]{majewski2017apache}, a part of the Sloan Digital Sky Survey III (SDSS-III) \citep{eisenstein2011sdssiii} and APOGEE-2, a part of SDSS-IV \citep{blanton2017sloan}, has yielded a rich database of $R \sim 22,500$ and $S/N > 100$ spectra and derived data products (stellar parameters and multi-element chemical abundances) for nearly nearly 3/4 million stars across the MW and its satellite system.  APOGEE-2 had as a specific science driver the exploration of the MCs, which motivated placement of a second, APOGEE-S spectrograph \citep{wilson19spectro} in the Southern Hemisphere.
This work is part of a series of three papers on the Magellanic Clouds and their chemistry, as revealed by 
APOGEE
data. Povick et al. 2023a (submitted; hereafter Paper I), presents a method to calculate ages of individual RGB stars using high-resolution spectroscopy, multi-band photometry, and isochrones and applies it to the APOGEE-2S sample of the LMC. In Povick et al. 2023b (submitted; hereafter Paper II), the abundance gradients of the LMC and their temporal evolution are determined. The focus of this third paper in the series is the abundance gradients of the SMC

In this paper, after discussion of 
the APOGEE and Gaia data (Section \ref{sec:smc_data}), 
Section \ref{sec:smc_distage} outlines how we estimate distances and ages for the SMC stars.
Section \ref{sec:smc_gradmethod} outlines abundance trend calculations, while their results are presented in Section \ref{sec:smc_results}. From there, we discuss the implications of these results in Section \ref{sec:smc_discussion}.  Finally, concluding remarks are given in Section \ref{sec:smc_summary}.


\begin{figure}
    \centering
    \includegraphics[scale=0.43]{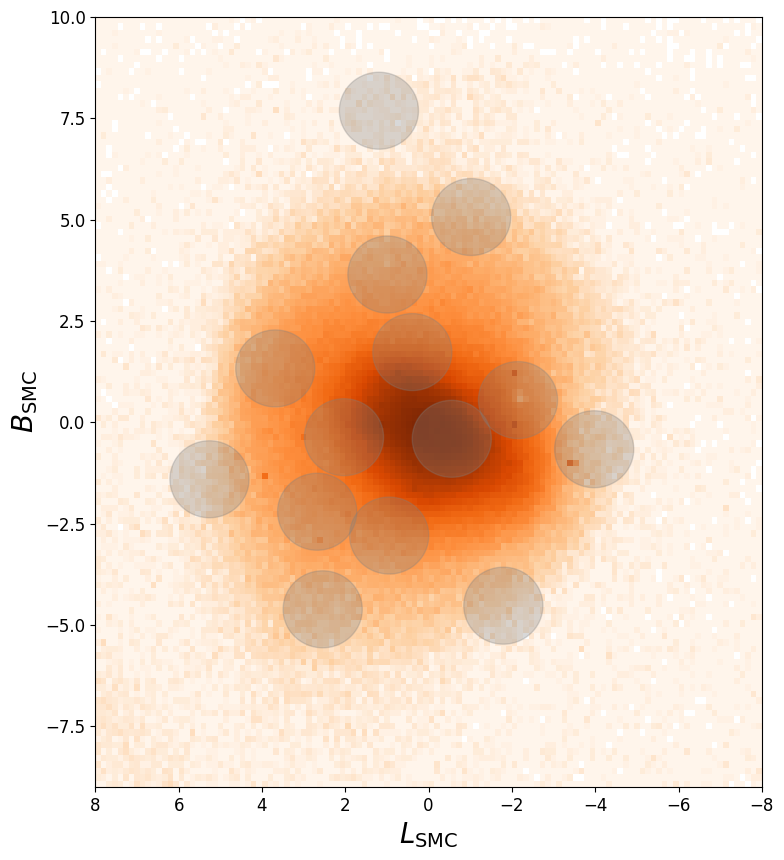}
    \caption{Stellar density map of the Gaia DR3 RGB with an SMC-like proper motions in an SMC-centric coordinate system in degrees. The fourteen APOGEE-2S SMC fields are highlighted in gray.}
    \label{fig:smc_map}
\end{figure}

\section{Data}
\label{sec:smc_data}

For our study of the SMC, we not only use the high-resolution, high-S/N spectroscopic data from APOGEE, but  also
multi-band photometry and astrometry from Gaia.


\subsection{APOGEE Data}
\label{subsec:smc_apogee_data}


Originally, the APOGEE survey was a single \textit{H}-band spectrograph \citep{wilson19spectro} linked to the 2.5-m Sloan Telescope at Apache Point Observatory \citep[APO,][]{gunn2006apo} in New Mexico. With the most recent iteration of APOGEE (APOGEE-2), a second APOGEE spectrograph was installed at the Las Campanas Observatory \citep[LCO,][]{bowen1973optical} in Chile, making it possible for APOGEE to obtain data in the Southern Hemisphere. 
The APOGEE data on the MCs were taken entirely as part of the second phase, APOGEE-2 program in 
SDSS-IV
\citep[][]{blanton2017sloan}.
All of the APOGEE-2 data are available in 
SDSS-IV data release 17 \citep[DR17,][]{Abdurro'uf2022ApJSdr17}.
Specific information about the targeting for APOGEE and APOGEE-2 can be found in \cite{zasowski13target} and \cite{zasowski17target}, while targeting information for the APOGEE-2 MC survey is in \citet{nidever20lazy}, which outlines specific selection criteria 
that minimize MW contamination in the high fidelity MC sample.



Stellar atmospheric parameters and chemical abundances are derived from the observed spectra in a two-step process. First, the observed spectra are fed into the APOGEE pipeline \citep{nidever15pipe} for basic reduction (1-D spectral extrctions from the 2-D images, wavelength calibration, flat-fielding, sky subtraction, etc.). Accurate stellar parameters are then calculated from the reduced spectra using the APOGEE Stellar Parameter and Chemical Abundance Pipeline \citep[ASPCAP,][]{holtzman15abund,perez16aspcap}. This second pipeline utilizes the \texttt{FERRE} \citep{allendeprieto2006} package to match the \teffe, $\log(g)$, v$_{\text{micro}}$, [M/H], [C/M], [N/M], [$\alpha$/M], and v$_{\text{macro}}$ to a library of synthetic spectra created with \texttt{synspec} \citep{hubeny21synspec}. The list of lines used to derive abundances for different elements by ASPCAP can be found in \cite{smith2021lines} and \cite{shetrone2015lines}. 
Elemental abundances that are measured by ASPCAP include C, C\textsc{i}, N, O, Na, Mg, Al, Si, S, K, Ca, Ti, Ti \textsc{ii}, V, Cr, Mn, Fe, Co, Ni, and Ce. For DR17, the process of deriving these abundances is nearly identical to that in previous data releases except for some details outlined in Holtzman et al. (in prep.).  More information on the APOGEE derivation of abundances for the s-process element Ce can be found in \cite{cunha2017adding}.



As with any measurements, the APOGEE results have both statistical uncertainties as well as some systematic errors.  To attempt to remove the latter, potential systematic biases in the raw APOGEE measurements are determined by inspecting stars in the solar neighborhood with solar metallicities. Element specific abundance zero point offsets are then derived for these solar metallicity stars and applied to 
the entire APOGEE database. The statistical uncertainties were calculated from the scatter seen in multiple visits to the same stars processed independently 
as functions of \teffe, [M/H], and SNR.\footnote{
The objective of APOGEE is to produce spectra with SNRs of 100 per half-resolution unit \citep[e.g.,][]{jonsson2020apogee}.} For more on the specifics of the offsets and uncertainties, please refer to \citep{nidever20lazy,jonsson2020apogee} and the discussion therein. For calculations used here,
these calibrated abundances from ASPCAP values are used unless otherwise specified.




APOGEE was able to observe 2062 SMC RGB stars 
out to $\sim$9 kpc in 14 different fields (see Fig.\ \ref{fig:smc_map}).
This relatively large radial coverage makes the APOGEE SMC survey well-suited for determining the abundance gradients in the galaxy.

\begin{figure}
    \centering
    \includegraphics[scale=0.3]{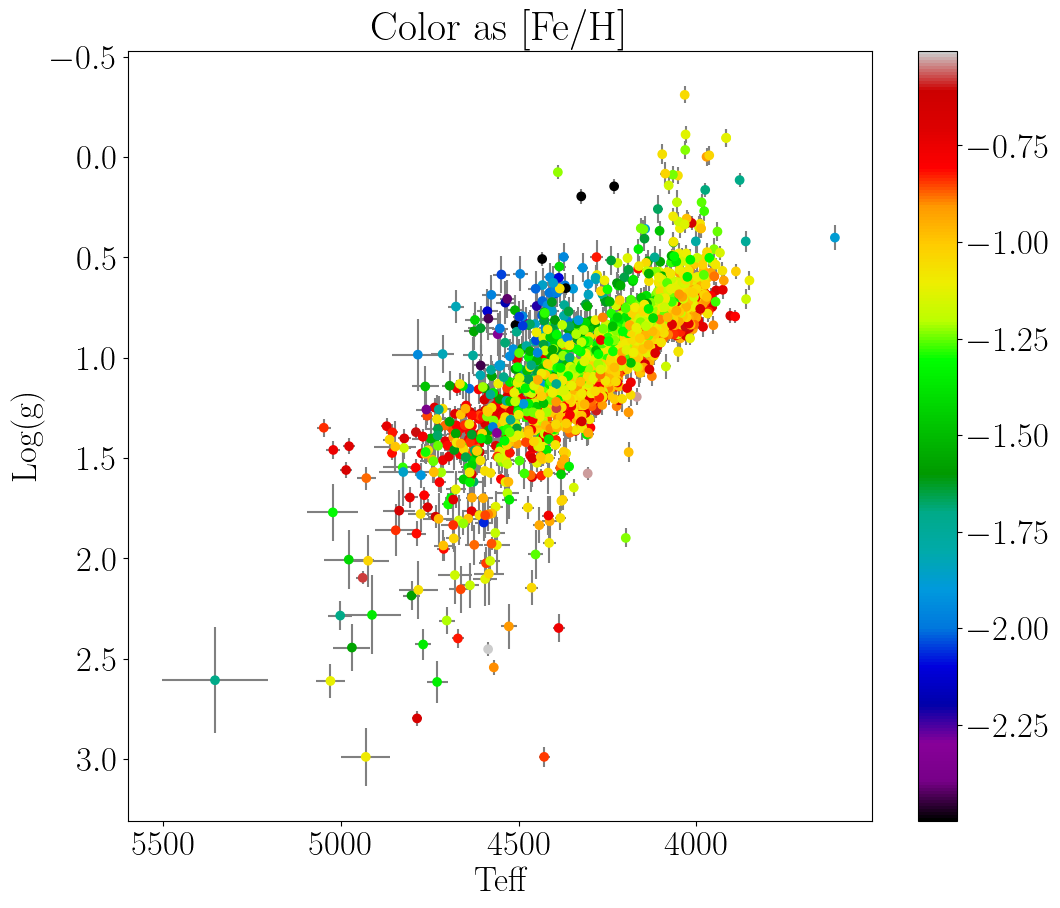}
    \caption{A Kiel diagram for APOGEE SMC RGB stars. The stars are color-coded by their metallicity. Error bars showing the \teff and \logg uncertainties are shown in gray.}
    \label{fig:smc_HR}
\end{figure}




\subsection{Gaia Data Release 3}

The Gaia mission \citep{gaia2016gaia,gaia2022dr3} is a massive all-sky photometric, astrometric and spectroscopic survey at optical wavelengths. The three Gaia photometric bands are ${BP}$ (blue), $G$ (green), and ${RP}$ (red).
The Gaia Archive stores all the publicly available data and can be accessed through the Gaia website\footnote{\url{https://gea.esac.esa.int/archive/}}. 

For this work, we use the Gaia red clump stars (RCs) in the APOGEE fields to derive ``proper motion distances'' (see Section \ref{ssec:smc_distances}). The first step was to query the Gaia Archive for the SMC stars using a slightly modified version of the selection used by \cite[][hereafter ``L21'']{luri2021mcs}. 
The largest deviation from the L21 criteria is using the center of the SMC from \cite{nidever2017smash} and some slightly more restrictive quality cuts. While it is unknown how many of these stars may be MW foreground contamination, it should be comparable to the contamination in the sample from L21. For the full ADQL query, see Appendix \ref{app:smc_gaia_adql}. 

Next, the RC stars are selected by using color-magnitude diagram criteria determined by eye:

\begin{enumerate}
    \item 0.875 $<$ ${BP}$ $-$ ${RP}$ $<$ 1.2
    \item 18.6 $<$ $G$ $<$ 19.6
\end{enumerate}

\noindent
Finally, for each APOGEE SMC field, all candidate RC stars within 0.95$^\circ$ of the field center are selected,
where 0.95$^\circ$ is the angular radius of the APOGEE plates. 



\section{Calculating Distances and Ages}
\label{sec:smc_distage}

In Paper I, we determined ages for individual LMC RGB stars using spectroscopically-determined stellar parameters, multi-band photometry, and isochrones.  In Paper II, we used these ages to determine the temporal evolution of LMC elemental abundances and its radial abundance gradients.  An important ingredient of our age-determination method is that we know the distances of the LMC RGB stars quite accurately because they lie predominantly within an inclined disk plane with observationally well-determined parameters.
This allowed us to calculate absolute magnitudes that can be compared to isochrone photometry to
constrain the star's age.
However, the SMC stars do not lie in an inclined disk plane, but, instead, have a complex, somewhat spheroidal geometry \citep[e.g.,][]{zaritsky2000morphology} with a large line-of-sight depth on the eastern side.  Thus, another technique to obtain the distances to SMC stars is needed to implement 
our age-determination method.




\subsection{SMC Distances}
\label{ssec:smc_distances}



The kinematics of a star can be used to estimate its distance by way of its proper motion, which has a 
linear dependence on distance.  The 
space 
velocity of an object is 
given by
\begin{equation}\label{equ:velocity}
    \begin{array}{l}
        v^2 = v_r^2 + v_t^2
        = v_r^2 + (4.74\mu d)^2,
    \end{array}
\end{equation}
\noindent
where $v_r$ is the radial velocity, $v_t$ is the tangential velocity, $\mu$ is the magnitude of the proper motion in units of mas yr$^{-1}$, and $d$ is the distance to the object in units of kpc (with all quantities in a heliocentric reference frame).


For our APOGEE SMC RGB sample, the radial velocities are known from the APOGEE spectroscopy and the proper motions from Gaia astrometry.
Although the tangential velocity ($v_t$) of any RGB star is unknown, it can be estimated by using the Gaia RC stars.  These stars are standard candles, and, therefore, their distances can be accurately established simply from their photometry.  While the RC absolute magnitude is fairly constant in near-infrared (NIR) magnitudes \citep{Bovy2014}, it has a metallicity-dependence in the optical bands \citep{Girardi2016}.  Using PARSEC isochrones, we determine that the $G$-band absolute magnitude of the RC for the mean metallicity of the SMC ([Fe/H] $=-1.1$) is $G_{\rm RC,SMC} = -0.21$.
After determining the distance of each individual RC star, its tangential velocity can be calculated by using its distance and proper motion via $v_t = 4.74\mu d$. 
Next, we calculate the mean tangential velocity of all the Gaia RC stars in each APOGEE field separately.  
Finally, the distance of an individual RGB star can be calculated from its $\mu$ under the assumption that its tangential velocity is the same as the mean RC tangential velocity in its field, using
\begin{equation}\label{equ:distance}
   d_{\rm RGB} = \langle v_{t,{\rm RC}} \rangle / (4.74 \mu)
\end{equation}

We find that the mean distance for the SMC RGB stars
using this method is 66.21 kpc. An offset of 3.77 kpc is then applied so that the mean distance of the RGB stars matches the 62.44 kpc found by \cite{graczyk2020distance}. 

\begin{figure*}
    \centering
    \includegraphics[width=\textwidth]{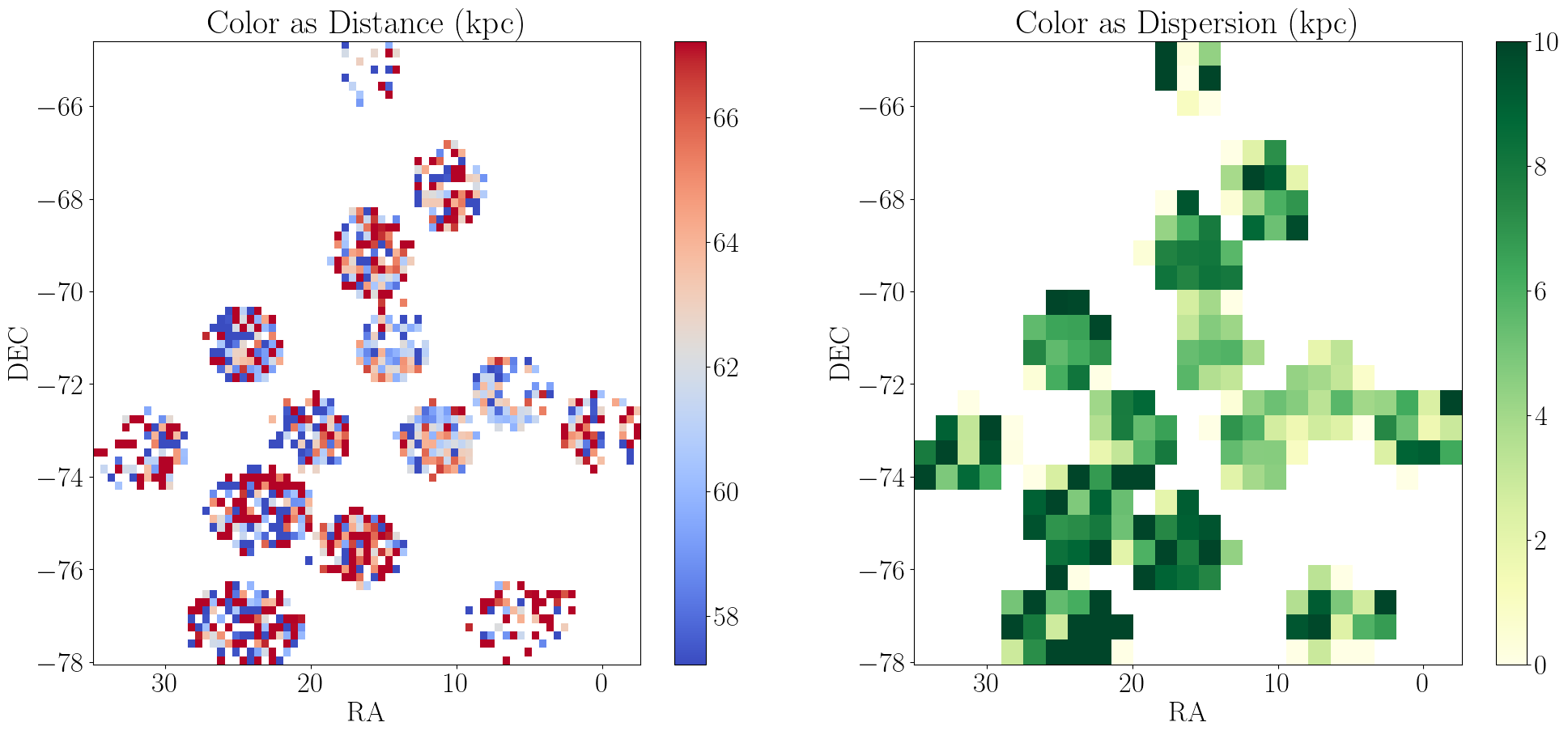} 
    \caption{({\em Left}) A map of the APOGEE SMC fields showing the mean distance in each spatial bin determined using our ``proper motion distance'' method. ({\em Right}) A dispersion map for the calculated distances. The larger line-of-sight depth is clearly visible on the eastern side of the galaxy.}
    \label{fig:smc_distance}
\end{figure*}

Naturally, the actual tangential velocity of any given RGB star will deviate from the field-averaged mean.  This will result in an error in our distance calculation.  We can estimate this bias by using the measured radial velocity dispersion and assuming that the tangential velocity dispersion is similar.  The average $v_r$ dispersion of the 14 SMC fields is 22.7 \kmse.  Using propagation of errors, this $v_t$ dispersion translates to 3.25 kpc in distance.  At an average SMC distance of 62.44 kpc, this amounts to a 5.2\% uncertainty, which is quite good.  The Gaia proper motions of the relatively bright individual SMC RGB stars have a fractional uncertainty of 4.2\% which also translates to a fractional uncertainty in distance of 4.2\% or 2.6 kpc.  Combining these two sources of uncertainty in quadrature gives us an overall distance uncertainty of 4.16 kpc or 6.7\% on a star-by-star basis. There are also systematic errors, as evidenced by the 3.77 kpc offset applied.
Figure \ref{fig:smc_distance} shows a map of the APOGEE SMC distances determined using this method showing a higher distance dispersion on the eastern side, as previously reported.

\subsection{SMC Ages}
\label{sssec:smc_ages}

\begin{figure}
    \centering
    \includegraphics[width=0.475\textwidth]{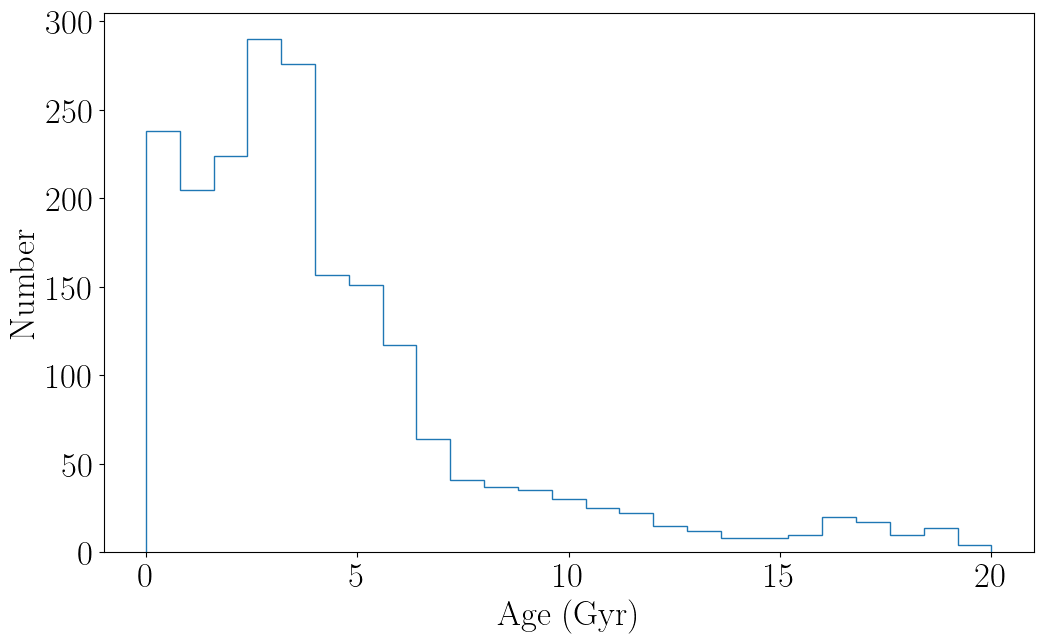} 
    \caption{The distribution of calculated ages for the SMC stars. A large portion of the stars formed within the last $\sim$7.5 Gyr in the SMC with a peak $\sim$3--4 Gyr ago. This star formation history is very reminiscent to that seen for the LMC (see Figure 19 of Paper I).
    In addition, the feature at $\sim$5 Gyr corresponds to one of the peaks in the SFH from \protect\cite{rubele2018sfh}.}
    \label{fig:smc_age_dist}
\end{figure}

Using the age method outlined in Paper I and the distances from the previous section, ages can be derived for the SMC stars. We find that the distribution of our derived SMC ages is shown in Figure \ref{fig:smc_age_dist} and looks quite similar to the LMC age distribution shown in Figure 19 of Paper I. In the age distribution of the SMC there appears to be a feature around 5 Gyr suggesting a steadily rising
star formation before the largest SF peak. The peak in the SFR centered at 5 Gyr and spanning 3.98--6.31 Gyr ago in \cite{rubele2018sfh} overlaps with both the 5 Gyr feature and the global peak seen in Figure \ref{fig:smc_age_dist} herein and is the most likely cause of these two.


The 6.7\% or 4.16 kpc distance uncertainty that we calculated in Section \ref{ssec:smc_distances} does have an impact on the ages.  Propagating the uncertainty forward shows tat the distance uncertainty 
produces an average age uncertainty up to 24\% of the calculated age. 

\section{Abundance Gradient Calculation}
\label{sec:smc_gradmethod}

Here we discuss the details of how the SMC radial abundance gradients are determined for each of the abundance ratios measured by the APOGEE observations.  However, how we define radial distances for SMC stars must be handled differently than how we did so for LMC stars in Paper II.

\subsection{SMC Radius}\label{sssec:smc_radius}


Because the SMC does not have a well-shaped stellar disk, but rather a more spheroidal shape, the best way to project the stellar sky positions to a specific galactic position is not obvious.
Therefore
we have elected to derive our abundance gradients simply in terms of each star's distance 
from the center of the SMC as projected on the plane of the sky.
Thus, we determine the SMC radius of each star 
by converting the 
celestial coordinates ($\alpha$, $\delta$) into angular coordinates ($\rho$, $\phi$), where $\rho$ is the angular radius from the center of the SMC and $\phi$ is the position angle. The angular coordinates in terms of $\alpha$ and $\delta$ are 
\begin{equation} \label{equ:angcoor}
    \begin{array}{l}
    \cos{\rho} = \cos{\delta_0}\cos{\delta}\cos{(\alpha-\alpha_0)} + \sin{\delta_0}\sin{\delta}, \\
    \sin{\rho}\cos{\phi} = -\cos{\delta}\sin{(\alpha-\alpha_0)}, \\
    \sin{\rho}\sin{\phi} = \cos{\delta_0}\sin{\delta} - \sin{\delta_0}\cos{\delta}\cos{(\alpha-\alpha_0)},
    \end{array}
\end{equation}
where $(\alpha_0, \delta_0$) is the center of the SMC at (13.18$^\circ$, $-$72.83$^\circ$) from \cite{nidever2017smash}. 
Converting the angular radius, $\rho$, to a linear radial is done by
\begin{equation}\label{equ:radius}
    R = D\tan{\rho}
\end{equation}
\noindent
where $D$ is the distance to the center of the SMC, adopted as
62.44 $\pm$ 0.44 kpc from \cite{graczyk2020distance}.

\subsection{Determining Abundance Gradients with MCMC}
\label{ssec:smc_mcmc}

Radial abundance gradients are determined by fitting abundance trends with radius in a galaxy. The convention is to use a simple linear model to accomplish this, and, therefore, the functional form of the model adopted here 
is
\begin{equation}\label{equ:linmodl}
    [\text{X/Fe}] = \nabla_{R}R+[\text{X/Fe}]_0,
\end{equation}
\noindent
where [X/Fe] is the abundance ratio normalized to iron, $\nabla_R$ is the gradient of the line, $R$ is the radius (distance from the center of the SMC), and [X/Fe]$_0$ is the central abundance ratio. 

Measuring chemical abundances with respect to hydrogen as the fiducial element is taken as the ``absolute'' abundance for other elements.
However, 
to monitor temporal variations in the contributions of chemical yields from different nucleosynthetic processes in the overall chemical evolution of a galaxy it is useful to normalize out the overall increase in metallicity with time.  To do so, it is common to look at variations in abundance ratios from solar-scaled values by normalizing against [Fe/H] --- 
that is, [X/Fe] = [X/H] $-$ [Fe/H]. 
However, 
\citet{weinberg2019chemical} have argued that 
magnesium is a better fiducial element for abundance ratios
because it is released into the interstellar medium almost exclusively through Type II SNe \citep{andrews2017inflow}, whereas iron has contributions from both SNe II and SNe Ia, where the latter are subject to delays related to the time required for the creation of white dwarf binaries.
Thus, we determine gradients for [X/Fe], [X/H], and [X/Mg], 
as well as for the key diagnostic ratios [C/N] and [$\alpha_\text{h}$/$\alpha_\text{ex}$]. The [C/N] of a star has been shown to be a good indicator of age for
metal rich giant stars \citep{ness2016spectroscopic}. On the other hand, [$\alpha_\text{h}$/$\alpha_\text{ex}$] is ratio of the typically lighter hydrostatic $\alpha$-elements to the heavier explosive $\alpha$-elements (see Section \ref{ssec:smc_alpha_over}).

The gradients and central abundances in Equation \ref{equ:linmodl} are found using maximum likelihood estimation with the likelihood function:
\begin{equation} 
\label{equ:lnL}
    \ln{\mathcal{L}} = \frac{1}{2}\bigg[ \sum_{i} \frac{([\text{X/Fe}]_i - \nabla_RR_i - [\text{X/Fe}]_0)^2}{\sigma_i^2} - \ln{2\pi \sigma_i^2} \bigg],
\end{equation}
\noindent
where the distance of each individual star from the center of the SMC is $R_i$, the abundance of the star is [X/Fe]$_i$, and the error in the abundance is $\sigma_i$.

For added robustness when fitting the abundance gradients, all stars with S/N $<$ 100 are removed, the stars are binned in radius, and a model is fit to the median of the binned values with weighting using the median absolute deviation (MAD) from the median. After this fit, any outlier stars further than 3$\times \sigma$ are rejected and the model is refit. This is done because there is large abundance scatter at each radius especially in the outer galaxy.  Uncertainties in the gradients and central abundances are found by implementing a Markov chain Monte Carlo (MCMC) analysis using the \texttt{emcee} \citep{foreman2013emcee} python package. This package is built upon the algorithms presented in \citet{goodman2010ensemble}. 

\subsection{Stellar Age Binning}
\label{ssec:smc_age}

As mentioned above, we use the calibrated spectroscopic parameters (\teffe, \logge, and [Fe/H]) and photometry as well as the distances calculated from Section \ref{ssec:smc_distances} to determine ages using the method from Paper I.
To simplify comparison of the evolution of abundance gradients between the LMC and SMC, the same age bins from Povick et al.\ (2023b, submitted) are used here. The particular age ranges for each bin can be found in Table \ref{tab:smc_age_bin}.



\begin{table}
	\centering
	\caption{The age bins used throughout this work, which are the same as used for our analysis of the LMC in Povick et al. (2023b). $N$ is the total number of stars in each bin and $N_{\text{SNR}>100}$ is the number of stars in each bin that have an SNR over 100. For a few stars ages could not be derived;
this leads to a discrepancy between
the sum of the $N$ column values and 
the previously stated sample size in Section \ref{subsec:smc_apogee_data}. 
}
	\label{tab:smc_age_bin}
	\begin{tabular}{cccc} 
		\hline
		Bin & Age Range & $N$ & $N_{\text{SNR}>100}$ \\
         & (Gyr) & & \\
		\hline
		1 & t $\leq$ 2.23 & 613 & 213 \\
		2 & 2.23 $<$ t $\leq$ 3.66 & 503 & 172 \\
		3 & 3.66 $<$ t $\leq$ 5.58 & 422 & 164 \\
            4 & 5.58 $<$ t $\leq$ 8.36 & 247 & 83 \\
            5 & 8.36 $<$ t & 263 & 76 \\
		\hline
	\end{tabular}
\end{table}

\subsection{Age-[X/Fe] Trends}
\label{ssec:smc_age_xfe_trend}

In addition to radial abundance trends, we also explore the 
age-abundance ratio
trends 
for the [X/Fe] abundance ratio, as well as for [C/N], [$\alpha_\text{h}$/Fe], [$\alpha_\text{ex}$/Fe] and [$\alpha_\text{h}$/$\alpha_\text{ex}$]. Stars were divided by their SMC position angle into east and west regions. The eastern part of the SMC is known to have a larger line-of-sight depth compared to that in the western part of the galaxy \citep{Nidever2013smc}. On the other hand, the western part of the SMC contains the counter-bridge \citep{Diaz2012} and the west halo \citep{dias2016westhalo}. 
By dividing the sample along an east-west line, these structures are better isolated and 
probed. In addition, each of the angular bins was further divided into two radial annular bins. Stars were classified as either inner ($R < 3.3$ kpc) or outer ($3.3 \leq R$ kpc) SMC. This radial cutoff was chosen because the inner SMC is morphologically more elliptical for $R < 3^{\circ}$ compared to the outer portion of the galaxy \citep{nidever2011discovery}, which corresponds to 3.3 kpc.




The age-abundance ratio
trends are found by splitting the stars into different age bins and finding the abundance median and MAD scatter of each bin. Note that the age bins here are not the same as the ones used for exploring the evolution of the radial abundance gradients. The optimal number of bins was found to be 15 after trying different schemes. This number minimized the amount of bins with low numbers while also not washing out the general trend.



\section{Results}
\label{sec:smc_results}

Below, we present the main results of the radial gradients calculated for 24  abundance ratios and age-abundance ratio trends.

\subsection{The SMC Radial Abundance Ratio Gradients}
\label{ssec:smc_over_grad}

The SMC radial abundance gradients are discussed separately in element groups. Generally, abundance ratio gradients for lighter species are presented before heavier ones.

\subsubsection{Carbon \& Nitrogen}
\label{ssec:smc_cn_over}



This work looks at the individual gradients for carbon and nitrogen, the composite C+N gradients, and the [C/N] abundance ratio. The largest contributor to measured carbon and nitrogen elements is intermediate mass stars that undergo Type II SNe explosions \citep{kobayashi2006chemical,ventura2013agb}. Dredge-up (which is not corrected for here) does affect the observed C and N values of a star, but the observed C+N abundances do not change much \citep{gratton2000mixing}.


The fit of the radial trends to the [X/Fe] and [C/N] abundance ratios can be found in Figure \ref{fig:smc_cn_radxfe}. The carbon gradients of [C/Fe], [C/H], and [C/Mg] all are steep, especially the [C/H] gradient, with a value $-$0.0929 $\pm$ 0.0077 dex/kpc (see Table \ref{tab:smc_cn_over_grad}). On the other hand, in comparison, the nitrogen gradients are very shallow. The largest nitrogen gradient is $-$0.0427 $\pm$ 0.0046 dex/kpc for [N/H]. The C+N gradients fall in between the carbon and nitrogen gradients, which is expected because the C+N abundances are calculated using a weighted average of the individual carbon and nitrogen abundances. Lastly, the [C/N] gradient is low, relatively speaking, with a value of $-$0.0203 $\pm$ 0.0040 dex/kpc.


\begin{table}
	\centering
	\caption{Radial gradients for the abundant elements carbon and nitrogen.}
	\label{tab:smc_cn_over_grad}
	\begin{tabular}{cccc}
		\hline
		Element & $\nabla_R$ [X/Fe] & $\nabla_R$ [X/H] & $\nabla_R$ [X/Mg]\\
                    & (dex/kpc) & (dex/kpc) & (dex/kpc) \\
            \hline
            C & $-$0.0267 $\pm$ 0.0029 & $-$0.0929 $\pm$ 0.0077 & $-$0.0196 $\pm$ 0.0033 \\
            N & $-$0.0048 $\pm$ 0.0018 & $-$0.0427 $\pm$ 0.0046 & $-$0.0008 $\pm$ 0.0031 \\
            C+N & $-$0.0117 $\pm$ 0.0017 & $-$0.0676 $\pm$ 0.0059 & $-$0.0059 $\pm$ 0.0022 \\
            \hline
            {[C/N]} & $-$0.0203 $\pm$ 0.0040 & ... & ... \\
            \hline
	\end{tabular}
\end{table}

\subsubsection{\texorpdfstring{$\alpha$}\,-Elements}
\label{ssec:smc_alpha_over}

There are a number of $\alpha$ elements measured by APOGEE including O, Mg, Si, S, Ca, Ti, and overall $\alpha$. Enrichment of $\alpha$-elements in the interstellar medium (ISM) primarily occurs though Type II SNe \citep{nomoto2013nucleosynthesis,weinberg2019chemical}. From individual $\alpha$-elements, it is possible to calculate the average hydrostatic $\alpha$-abundance ([$\alpha_h$/Fe]=([\text{O/Fe}]+[\text{Mg/Fe}])/2), the average explosive $\alpha$-element abundance ([$\alpha_\text{ex}$/Fe] = ([\text{Si/Fe}]+[\text{Ca/Fe}]+[\text{Ti/Fe}])/3), and the ratio of the average hydrostatic $\alpha$-abundance to the average explosive $\alpha$-abundance ([$\alpha_\text{h}/\alpha_\text{ex}$] = [$\alpha_h/\text{Fe}$]-[$\alpha_\text{ex}/\text{Fe}$]). Hydrostatic $\alpha$-elements are so named because they are created from hydrostatic burning inside of a star while the explosive $\alpha$-elements are created during Type II SNe itself \citep{carlin2018chemical}.

Regardless of the fiducial element, all of the $\alpha$-abundances have a wide range of values, as seen in Table \ref{tab:smc_alpha_over_grad}. It is immediately clear that the [X/Fe] gradients are almost all positive, which is the opposite of what is seen in the [X/H] and [X/Mg] gradients. The only negative [X/Fe] gradients are for [Ti/Fe], with a value of $-$0.0115 $\pm$ 0.0018 dex/kpc, and [$\alpha_\text{ex}$/Fe], with a value of $-$0.0001 $\pm$ 0.0011 dex/kpc. The [$\alpha_\text{ex}$/Fe] is probably negative because of the the [Ti/Fe] values.
The average of the [Si/Fe], [Ca/Fe], and [Ti/Fe] gradients is negative due to the steep negative gradient of [Ti/Fe], so it would be expected that the [$\alpha_\text{ex}$/Fe] would also be negative. The [X/Fe] and [$\alpha_\text{h}/\alpha_\text{ex}$] $\alpha$-element trends are shown in Figures \ref{fig:smc_ind_alpha_radxfe} and \ref{fig:smc_comb_alpha_radxfe}, where the global trends are shown in black.

\begin{table}
	\centering
	\caption{Table of radial abundance gradients for the $\alpha$-elements.}
	\label{tab:smc_alpha_over_grad}
	\begin{tabular}{cccc}
		\hline
		Element & $\nabla_R$ [X/Fe] & $\nabla_R$ [X/H] & $\nabla_R$ [X/Mg]\\
                    & (dex/kpc) & (dex/kpc) & (dex/kpc) \\
    		\hline
                O & 0.0024 $\pm$ 0.0013 & $-$0.0635 $\pm$ 0.0054 & 0.0029 $\pm$ 0.0010 \\
                Mg & 0.0074 $\pm$ 0.0019 & $-$0.0643 $\pm$ 0.0053 & ... \\
                Si & 0.0067 $\pm$ 0.0016 & $-$0.0512 $\pm$ 0.00482 & $-$0.0029 $\pm$ 0.0015 \\
                S & 0.0172 $\pm$ 0.0090 & $-$0.0502 $\pm$ 0.0094 & 0.0139 $\pm$ 0.0084 \\
                Ca & 0.0001 $\pm$ 0.0013 & $-$0.0668 $\pm$ 0.005 & $-$0.0072 $\pm$ 0.0019 \\
                Ti & $-$0.0115 $\pm$ 0.0018 & $-$0.0677 $\pm$ 0.0057 & $-$0.0075 $\pm$ 0.0022 \\
                $\alpha$ & 0.0004 $\pm$ 0.0009 & $-$0.0681 $\pm$ 0.0051 & $-$0.0023 $\pm$ 0.0011 \\
                $\alpha_\text{h}$ & 0.0051 $\pm$ 0.0017 & $-$0.0648 $\pm$ 0.0053 & 0.0014 $\pm$ 0.0005 \\
                $\alpha_\text{ex}$ & $-$0.0001 $\pm$ 0.0011 & $-$0.0647 $\pm$ 0.0055 & $-$0.0044 $\pm$ 0.0017 \\
                \hline
                {[$\alpha_\text{h}$/$\alpha_\text{ex}$]} & 0.0040 $\pm$ 0.0011 & ... & ... \\
    		\hline
	\end{tabular}
\end{table}

\subsubsection{Odd-Z Elements}
\label{ssec:smc_oddz_over}




The 
odd-Z elements
are produced mostly in Type II SNe with a slight metallicity dependence \citep{nomoto2013nucleosynthesis,weinberg2019chemical}. Regardless of the fiducial element, the K gradients fall between the Na and Al gradients (see Table \ref{tab:smc_oddz_over_grad}). The overall odd-Z abundance trends are shown as black lines in Figure \ref{fig:smc_oddz_radxfe}.

\begin{table}
	\centering
	\caption{Table of radial abundance gradients for the odd-Z elements.}
	\label{tab:smc_oddz_over_grad}
	\begin{tabular}{cccc}
		\hline
		      Element & $\nabla_R$ [X/Fe] & $\nabla_R$ [X/H] & $\nabla_R$ [X/Mg]\\
                    & (dex/kpc) & (dex/kpc) & (dex/kpc) \\
    		\hline
    		Na & 0.0214 $\pm$ 0.0084 & $-$0.0337 $\pm$ 0.0086 & 0.0140 $\pm$ 0.0087 \\
                Al & $-$0.0123 $\pm$ 0.0028 & $-$0.0713 $\pm$ 0.0066 & $-$0.0159 $\pm$ 0.0018 \\
                K & 0.0061 $\pm$ 0.0029 & $-$0.0541 $\pm$ 0.0049 & 0.0044 $\pm$ 0.0026 \\
    		\hline
	\end{tabular}
\end{table}

\subsubsection{Iron Peak Elements}
\label{ssec:smc_ironpeak_over}

An important group of elements are those in  the iron peak, which includes V, Cr, Mn, Fe, Co, and Ni. These elements are produced through Type Ia SNe, although there is a significant contribution from Type II SNe as well \citep{kobayashi2006chemical}. 

The iron abundance, or metallicity, is a very important property of a star as it is essentially a measure of 
its overall heavy element content. The [Fe/H] does not stand out among the [X/H] gradients, even with its value of $-$0.0546 $\pm$ 0.0043 dex/kpc, but the [Fe/Mg] gradient is the steepest of those of the iron peak [X/Mg] elements at $-$0.1230 $\pm$ 0.0097 dex/kpc (see Table \ref{tab:smc_iron_over_grad}). As for the other gradients, there does not seem to be any discernible pattern with atomic number. The steepest [X/Fe] gradient belongs to [V/Fe], with a value of $-$0.0237 $\pm$ 0.0107 dex/kpc. At $-$0.0339 $\pm$ 0.0107 dex/kpc, vanadium also has the steepest non-iron [X/Mg] gradient. The shallowest iron peak gradient is different for each fiducial element. The radial trends for the iron peak element ratios [X/Fe] and [Fe/H] can be found in Figure \ref{fig:smc_ironpeak_radxfe}.


\begin{table}
	\centering
	\caption{Table of radial abundance gradients for the iron peak elements.}
	\label{tab:smc_iron_over_grad}
	\begin{tabular}{cccc}
		\hline
		      Element & $\nabla_R$ [X/Fe] & $\nabla_R$ [X/H] & $\nabla_R$ [X/Mg]\\
                    & (dex/kpc) & (dex/kpc) & (dex/kpc) \\
    		\hline
                V & $-$0.0237 $\pm$ 0.0107 & $-$0.0633 $\pm$ 0.0101 & $-$0.0339 $\pm$ 0.0107 \\
                Cr & $-$0.0059 $\pm$ 0.0037 & $-$0.0551 $\pm$ 0.0064 & $-$0.009 $\pm$ 0.0045 \\
                Mn & $-$0.013 $\pm$ 0.0014 & $-$0.0784 $\pm$ 0.0049 & $-$0.0146 $\pm$ 0.0033 \\
                Fe & ... & $-$0.0546 $\pm$ 0.0043 & $-$0.1230 $\pm$ 0.0097 \\
                Co & $-$0.0081 $\pm$ 0.0033 & $-$0.0653 $\pm$ 0.0061 & $-$0.0100 $\pm$ 0.0034  \\
                Ni & $-$0.0028 $\pm$ 0.0006 & $-$0.0574 $\pm$ 0.0048 & $-$0.0005 $\pm$ 0.0021 \\
    		\hline
	\end{tabular}
\end{table}

\subsubsection{The Neutron Capture Element Cerium}
\label{ssec:smc_neutron_over}




One of the more massive elements with a measured abundance from APOGEE is cerium. It is an element created through both the 
\textit{r}- and \textit{s}- neutron-capture processes \citep{prantzos2020srprocess}. It is be a difficult abundance for APOGEE to measure, so while gradients have been calculated for [Ce/Fe], [Ce/H], and [Ce/Mg], it is best not to make any strong conclusions due to the larger uncertainties. The [Ce/Fe] and [Ce/Mg] gradients have values of $-$0.0596 $\pm$ 0.0021 and $-$0.0599 $\pm$ 0.0039, respectively. The [Ce/H] abundance is very large at $-$0.1115 $\pm$ 0.0094 dex/kpc. The [Ce/Fe] abundance trend can be seen in Figure \ref{fig:smc_sr_radxfe}.

\begin{table}
	\centering
	\caption{Table of radial abundance gradients for cerium.}
	\label{tab:smc_cerium_over_grad}
	\begin{tabular}{cccc}
		\hline
		      Element & $\nabla_R$ [X/Fe] & $\nabla_R$ [X/H] & $\nabla_R$ [X/Mg]\\
                    & (dex/kpc) & (dex/kpc) & (dex/kpc) \\
    		\hline
    		Ce & $-$0.0596 $\pm$ 0.0021 & $-$0.1115 $\pm$ 0.0094  & $-$0.0599 $\pm$ 0.0039 \\
    		\hline
	\end{tabular}
\end{table}

\subsection{The Evolution of the SMC Abundance Ratio Gradients}\label{ssec:smc_evolve}


As previously mentioned (Section \ref{ssec:smc_age}), we have calculated ages for the individual RGB stars in the SMC subset of the APOGEE catalog. Using age bins from Table \ref{tab:smc_age_bin}, the evolution of all the abundance gradients is determined.

\begin{figure*}
    \centering
    \includegraphics[width=\textwidth]{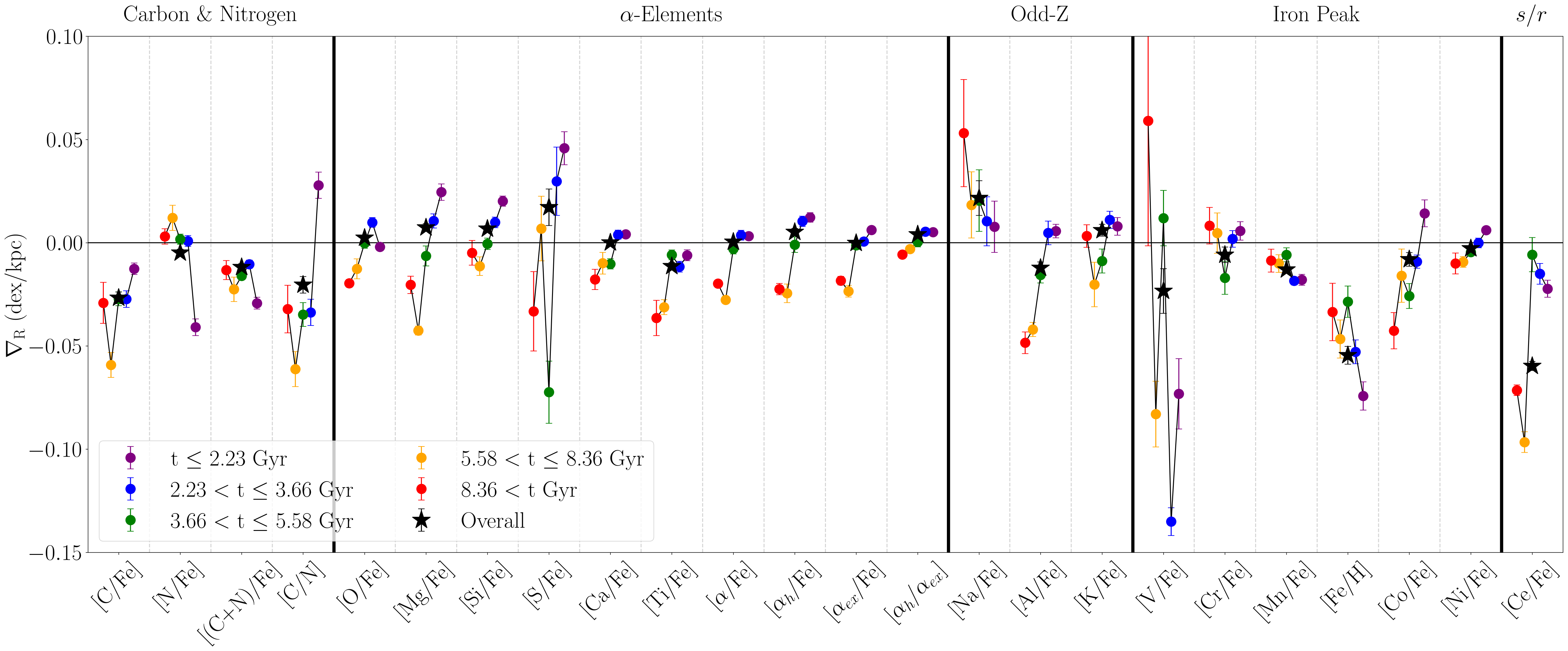}
    \caption{A figure showing each of the calculated gradients are shown for each element. The difference abundances have been broken up into different groups of like elements. The color of each dot corresponds to the different age bins that contain the same number of stars previously mentioned. The value of each gradient here can be seen in Table \ref{tab:smc_evolve_grad_fe}.}
    \label{fig:smc_all_grads_fe}
\end{figure*}

\begin{figure*}
    \centering
    \includegraphics[width=\textwidth]{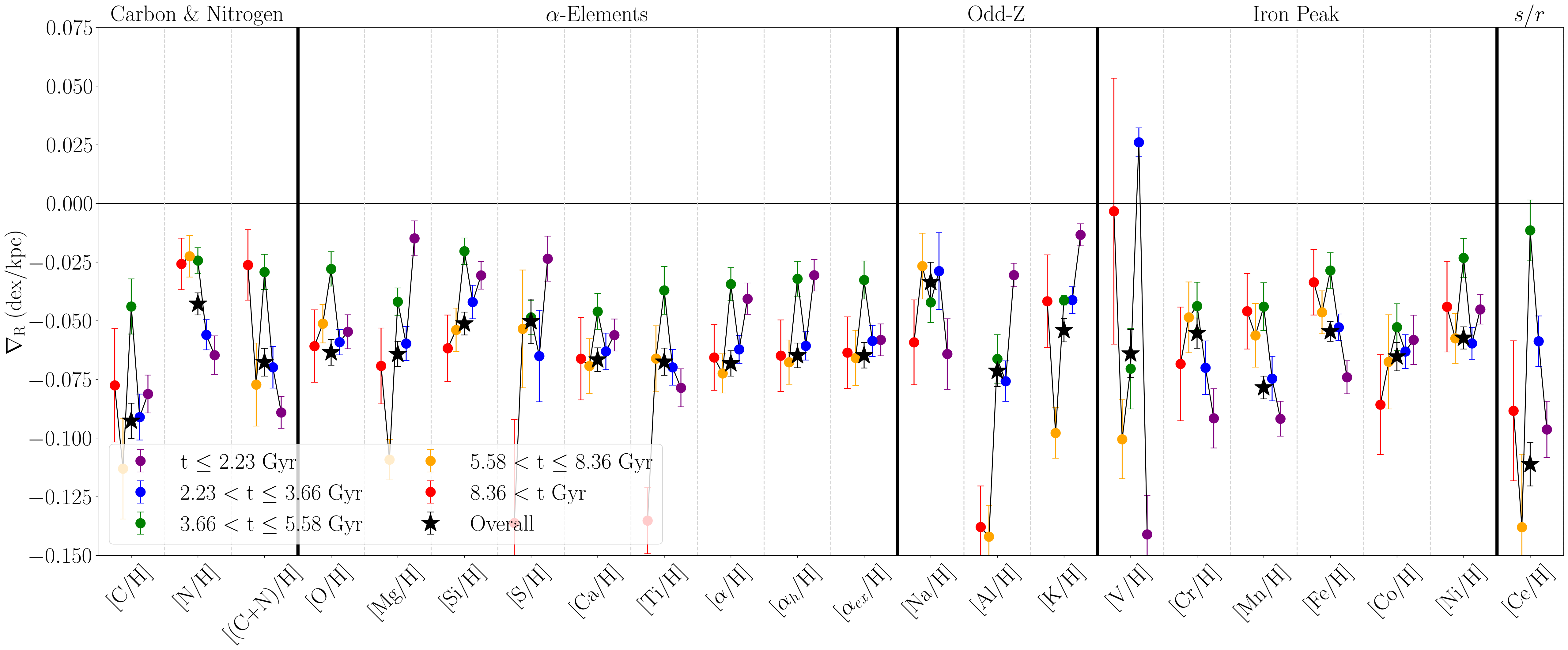}
    \caption{A figure showing each of the calculated gradients are shown for each element. The difference abundances have been broken up into different groups of like elements. The color of each dot corresponds to the different age bins that contain the same number of stars previously mentioned. The value of each gradient here can be seen in Table \ref{tab:smc_evolve_grad_h}.}
    \label{fig:smc_all_grads_h}
\end{figure*}

\begin{figure*}
    \centering
    \includegraphics[width=\textwidth]{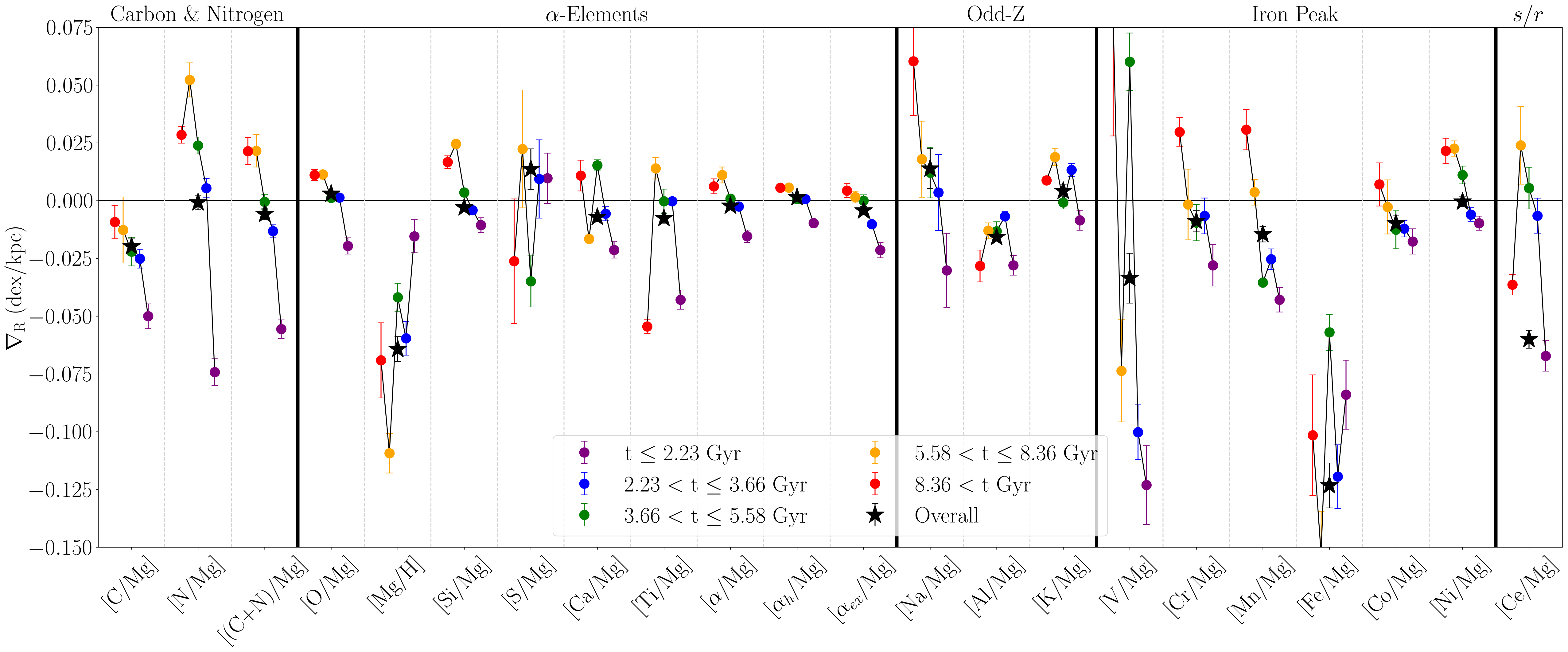}
    \caption{A figure showing each of the calculated gradients are shown for each element. The difference abundances have been broken up into different groups of like elements. The color of each dot corresponds to the different age bins that contain the same number of stars previously mentioned. The value of each gradient here can be seen in Table \ref{tab:smc_evolve_grad_mg}.}
    \label{fig:smc_all_grads_mg}
\end{figure*}

The evolutionary trends for all of the gradients with each fiducial can be seen in Figures \ref{fig:smc_all_grads_fe}, \ref{fig:smc_all_grads_h}, \ref{fig:smc_all_grads_mg}. For each abundance ratio the red points represent the gradient of the oldest stars while purple points show the gradient for the youngest stars, such that age/time runs left to right for each element ratio in the plots. Each of the nucleosynthetic groups of elements are grouped together, 
and for each group atomic number generally increases going to the right. A line in black has been added to connect the gradients to make the trends more obvious. The radial trends from which all of the gradients are derived are shown in Figures \ref{fig:smc_cn_radxfe}, \ref{fig:smc_ind_alpha_radxfe}, \ref{fig:smc_comb_alpha_radxfe}, \ref{fig:smc_oddz_radxfe}, \ref{fig:smc_ironpeak_radxfe}, and \ref{fig:smc_oddz_radxfe}. When fitting the radial trends, the number of stars falls off towards the outer galaxy leading to larger dispersions and sporadic behaviour when tracking the median abundance ratio value. Coupled with the different number of stars in each age bin and with the number of stars with good abundance values this can sometimes lead to a miss match between the fit radial trends and data especially for the oldest stars. This is precisely why outlier removal was used in Section \ref{ssec:smc_mcmc}.


\begin{table*}
	\centering
	\caption{Table of the [X/Fe] gradients for each of the age bins. Horizontal lines have been added marking the division of the previously defined groups of elements. In general descending down the table corresponds to an increase in atomic number. Each age bin has its own separate column.}
	\label{tab:smc_evolve_grad_fe}
	\begin{tabular}{cccccc}
        \hline
         & t $\leq$ 2.23 & 2.23 $<$ t $\leq$ 3.66 & 3.66 $<$ t $\leq$ 5.58 & 5.58 $<$ t $\leq$ 8.36 & 8.36 $<$ t \\
        Element & $\nabla_\text{R}$ & $\nabla_\text{R}$ & $\nabla_\text{R}$ & $\nabla_\text{R}$ & $\nabla_\text{R}$ \\
         & (dex/kpc) & (dex/kpc) & (dex/kpc) & (dex/kpc) & (dex/kpc) \\
        \hline 
        {[C/Fe]} & $-$0.0126 $\pm$ 0.0028 & $-$0.0273 $\pm$ 0.0040 & $-$0.0272 $\pm$ 0.0034 & $-$0.0592 $\pm$ 0.0060 & $-$0.0291 $\pm$ 0.0099 \\
        {[N/Fe]} & $-$0.0409 $\pm$ 0.0040 & 0.0008 $\pm$ 0.0026 & 0.0017 $\pm$ 0.0025 & 0.012 $\pm$ 0.0061 & 0.0030 $\pm$ 0.0037 \\
        {[(C+N)/Fe]} & $-$0.0293 $\pm$ 0.0028 & $-$0.0104 $\pm$ 0.0020 & $-$0.0161 $\pm$ 0.0008 & $-$0.0226 $\pm$ 0.0059 & $-$0.0133 $\pm$ 0.0046 \\
        {[C/N]} & 0.0278 $\pm$ 0.0063 & $-$0.0338 $\pm$ 0.0064 & $-$0.0348 $\pm$ 0.0058 & $-$0.0612 $\pm$ 0.0085 & $-$0.0321 $\pm$ 0.0115 \\
        \hline
        {[O/Fe]} & $-$0.0020 $\pm$ 0.0016 & 0.0098 $\pm$ 0.0024 & $-$0.0002 $\pm$ 0.0024 & $-$0.0126 $\pm$ 0.0048 & $-$0.0197 $\pm$ 0.0019 \\
        {[Mg/Fe]} & 0.0245 $\pm$ 0.0040 & 0.0106 $\pm$ 0.0034 & $-$0.0064 $\pm$ 0.0048 & $-$0.0426 $\pm$ 0.0022 & $-$0.0204 $\pm$ 0.0042 \\
        {[Si/Fe]} & 0.0201 $\pm$ 0.0025 & 0.0099 $\pm$ 0.0025 & $-$0.0005 $\pm$ 0.0027 & $-$0.0114 $\pm$ 0.0045 & $-$0.0049 $\pm$ 0.0060 \\
        {[S/Fe]} & 0.0458 $\pm$ 0.0080 & 0.0298 $\pm$ 0.0166 & $-$0.0724 $\pm$ 0.0150 & 0.0069 $\pm$ 0.0156 & $-$0.0332 $\pm$ 0.0192 \\
        {[Ca/Fe]} & 0.0041 $\pm$ 0.0017 & 0.0039 $\pm$ 0.0022 & $-$0.0102 $\pm$ 0.0025 & $-$0.0099 $\pm$ 0.0052 & $-$0.0178 $\pm$ 0.0049 \\
        {[Ti/Fe]} & $-$0.0061 $\pm$ 0.0026 & $-$0.0116 $\pm$ 0.0026 & $-$0.0059 $\pm$ 0.0024 & $-$0.0312 $\pm$ 0.0036 & $-$0.0364 $\pm$ 0.0086 \\
        {[$\alpha$/Fe]} & 0.0032 $\pm$ 0.0013 & 0.0036 $\pm$ 0.0024 & $-$0.0030 $\pm$ 0.0025 & $-$0.0277 $\pm$ 0.0012 & $-$0.0198 $\pm$ 0.0013 \\
        {[$\alpha_\text{h}$/Fe]} & 0.0122 $\pm$ 0.0023 & 0.0104 $\pm$ 0.0025 & $-$0.0010 $\pm$ 0.0037 & $-$0.0245 v 0.0045 & $-$0.0225 $\pm$ 0.0027 \\
        {[$\alpha_\text{ex}$/Fe]} & 0.0062 $\pm$ 0.0013 & 0.0006 $\pm$ 0.0019 & $-$0.0012 $\pm$ 0.0019 & $-$0.0235 $\pm$ 0.0028 & $-$0.0184 v 0.0021 \\
        {[$\alpha_\text{h}$/$\alpha_\text{ex}$]} & 0.0051 $\pm$ 0.0015 & 0.0053 $\pm$ 0.0016 & 0.0004 $\pm$ 0.0024 & $-$0.0031 $\pm$ 0.0019 & $-$0.0057 $\pm$ 0.0016 \\
        \hline
        {[Na/Fe]} & 0.0077 $\pm$ 0.0124 & 0.0104 $\pm$ 0.0119 & 0.0204 $\pm$ 0.0149 & 0.0183 $\pm$ 0.016 & 0.0531 $\pm$ 0.0259 \\
        {[Al/Fe]} & 0.0057 $\pm$ 0.0033 & 0.0048 $\pm$ 0.0057 & $-$0.0156 $\pm$ 0.0039 & $-$0.0420 $\pm$ 0.0034 & $-$0.0484 $\pm$ 0.0052 \\
        {[K/Fe]} & 0.0079 $\pm$ 0.0043 & 0.0111 $\pm$ 0.0042 & $-$0.0089 $\pm$ 0.0058 & $-$0.0202 $\pm$ 0.0108 & 0.0032 $\pm$ 0.0055 \\
        \hline
        {[V/Fe]} & $-$0.0732 $\pm$ 0.0170 & $-$0.1351 $\pm$ 0.0067 & 0.0119 $\pm$ 0.0134 & $-$0.0830 $\pm$ 0.0159 & 0.0591 $\pm$ 0.0606 \\
        {[Cr/Fe]} & 0.0057 $\pm$ 0.0045 & 0.0019 $\pm$ 0.0040 & $-$0.0170 $\pm$ 0.008 & 0.0047 $\pm$ 0.0098 & 0.0082 $\pm$ 0.0089 \\
        {[Mn/Fe]} & $-$0.0179 $\pm$ 0.0026 & $-$0.0185 $\pm$ 0.002 & $-$0.0059 $\pm$ 0.0034 & $-$0.0101 $\pm$ 0.0043 & $-$0.0087 $\pm$ 0.0055 \\
        {[Fe/H]} & $-$0.0742 $\pm$ 0.0068 & $-$0.0529 $\pm$ 0.0057 & $-$0.0286 $\pm$ 0.0076 & $-$0.0466 $\pm$ 0.0092 & $-$0.0335 $\pm$ 0.0139 \\
        {[Co/Fe]} & 0.0142 $\pm$ 0.0065 & $-$0.0091 $\pm$ 0.0032 & $-$0.0258 $\pm$ 0.006 & $-$0.0160 $\pm$ 0.0129 & $-$0.0426 $\pm$ 0.0088 \\
        {[Ni/Fe]} & 0.0061 $\pm$ 0.0008 & $-$0.0001 $\pm$ 0.0023 & $-$0.0045 $\pm$ 0.0010 & $-$0.0093 $\pm$ 0.0026 & $-$0.0101 $\pm$ 0.005 \\
        \hline
        {[Ce/Fe]} & $-$0.0223 $\pm$ 0.0041 & $-$0.0151 $\pm$ 0.0051 & $-$0.0058 $\pm$ 0.0083 & $-$0.0966 $\pm$ 0.005 & $-$0.0714 $\pm$ 0.0026 \\
    \end{tabular}
\end{table*}

\subsubsection{Carbon \& Nitrogen}
\label{ssec:smc_cn_evolve}

\begin{figure*}
    \centering
    \includegraphics[width=0.95\textwidth]{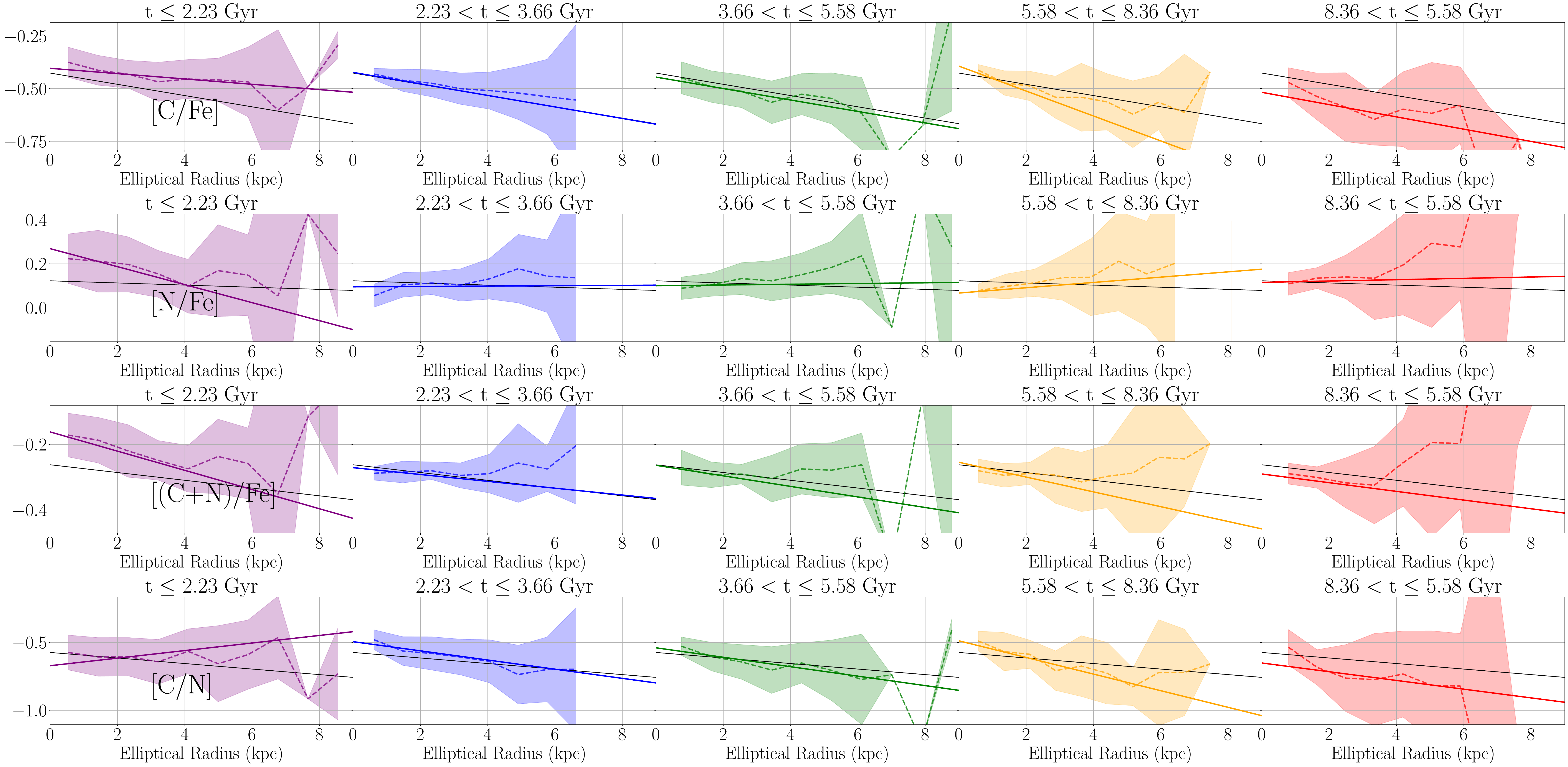}
    \caption{The radial abundance trends for the carbon and nitrogen abundances. For each row, age increases from left to right. The solid colored line is the fitted linear model, the dashed line follows the median abundance value as a function of radius, and the color band shows the dispersion around the median. The black line is the fit to all of the data without any age binning and serves as a reference fiducial. These plots show that each of the various abundance ratios increase over time, with the fitted trend lying below the reference fiducial for older age bins and above for younger age bins. The youngest [N/Fe] and [C/N] trends stand out by showing much larger gradients than the other age bins. Also the [C/N] gradient clearly inverts for the youngest age bin probably due to the rapid increase in [N/Fe] recently.}
    \label{fig:smc_cn_radxfe}
\end{figure*}

The carbon and nitrogen abundance gradients show different evolutionary trends depending on the particular fiducial element chosen. Starting first with Fe there is definite variation within the C and N gradients (see Figures \ref{fig:smc_all_grads_fe}). In the Figure, the [C/Fe] gradient has a U-shaped trend with a critical point for the stars aged 5.58--8.36 Gyr. Here U-shaped means that a gradients has experienced a flattening in the past, but then for more recent times has steepened. The individual trends may open up or down. For these stars there is an increase in the central [C/Fe] abundance, after which it remains constant around $-$0.4 dex (see Figure \ref{fig:smc_cn_radxfe}). The [N/Fe] gradient hovers around 0.0 dex/kpc, but for the stars with the youngest ages the gradient becomes quite negative, relatively speaking, with a value of $-$0.0409 $\pm$ 0.0040 dex/kpc. The means that recently the [N/Fe] has increased rapidly in the last few Gyr in the SMC center. The [(C+N/Fe)] gradient hovers around $-$0.0020 dex/kpc. The [C/N] gradient has a similar behaviour to [C/Fe], though it does become positive for stars younger than 2.23 Gyr. 


The [X/H] gradients for carbon and nitrogen all become more steeply negative for ages less than 5.58 Gyr (see Figure \ref{fig:smc_all_grads_h}). The [N/Fe] gradient has a U-shaped trend with an extremum point in the 5.58 $<$ t $\leq$ 8.36 Gyr age bin. As for the [C/H] and [(C+N)/H], if the oldest age bin is not considered, these also show a U-shaped trend, but the extremum point happens later than for [N/H], at around 3.66--5.58 Gyr ago. 

Lastly, for the carbon and nitrogen ratios with respect to magnesium, [C/Mg] shows a steepening trend with time while the [N/Mg] and [(C+N)/Mg] gradients start out positive for the oldest stars and then inverts, becoming negative (see Figure \ref{fig:smc_all_grads_mg}). The [N/Mg] gradient inversion lags the [(C+N)/Mg] gradient inversion, which happens in the 3.66--5.58 Gyr age bin.

\subsubsection{\texorpdfstring{$\alpha$}\,-Elements}
\label{ssec:smc_alpha_evolve}

The $\alpha$-element gradients have various evolutionary trends as seen in Figure \ref{fig:smc_all_grads_fe}. The [Mg/Fe] and [Si/Fe] gradients have parallel evolutionary patterns as evidenced by the same general shape of their gradient trends. The radial trends from Figure \ref{fig:smc_ind_alpha_radxfe} also show evolution in parallel for [Mg/Fe] and [Si/Fe], though the 5.58--8.36 gradient is much steeper for [Mg/Fe]. The relative evolution of [Ca/Fe] and [Ti/Fe] are similar in that they both tend to flatten out, but the gradients for stars younger than 3.66 Gyr tend to be slightly positive and the corresponding [Ti/Fe] gradients are negative (see Figure \ref{fig:smc_all_grads_fe}). Both [O/Fe] and [S/Fe] evolve differently from the rest of the individual $\alpha$-element gradients, but [O/Fe] is similar to [Ca/Fe]. The composite $\alpha$-element gradients of [$\alpha$/Fe], [$\alpha_\text{h}$/Fe], and [$\alpha_\text{ex}$/Fe] evolve much like [Ca/Fe] or [Ti/Fe] (see Figure \ref{fig:smc_all_grads_fe}). The radial trends of these also evolve similarly (see Figure \ref{fig:smc_comb_alpha_radxfe}). The [$\alpha_\text{h}$/$\alpha_\text{ex}$] gradient does not show much evolution and remains fairly flat for the whole history of the SMC.

Many of the evolutionary trends for the [X/H] gradients of the $\alpha$-elements have a notable maximum for stars with ages between 3.66 and 5.58 Gyr (see Figure \ref{fig:smc_all_grads_h}). While not necessarily a global extremum, these trends for the gradients show a significant change in the evolution trajectory. The [Mg/H], [Si/H], [$\alpha$/H], and [$\alpha_\text{h}$/H] gradients have this maximum, but then turn downturn with the 2.23$-$3.66 Gyr gradients becoming steeper. These gradients then flatten to around the 3.66$-$5.58 Gyr value or even more for the gradients of the youngest star. The [O/H], [Ca/H], [Ti/H] and [$\alpha_\text{ex}$/H] abundance ratios have global extrema for the 3.66$-$5.58 Gyr stars. The [Ca/H] evolutionary trend is also the only trend that shows a positive concave feature. The [S/H] gradient does not have a clear evolutionary trend and is fairly constant with time.




The individual $\alpha$-elements show quite a bit of variation in their evolution while the composite abundances are similar (see Figure \ref{fig:smc_all_grads_mg}). The evolution seen in [O/Mg] is comparable to the composite abundances. [S/Mg] does not have a clear evolutionary trend. The [Ti/Mg] gradients evolution has a U-shape with a critical point between 5.58--8.36 Gyr ago.

\subsubsection{Odd-Z Elements}
\label{ssec:smc_oddz_evolve}



Figure \ref{fig:smc_all_grads_fe} shows that all of the odd-Z elements have different evolutionary trends. The [Na/Fe] gradient starts as significantly positive with a value of $+$0.0531 $\pm$ 0.0259 dex/kpc and then flattens out to $+$0.0077 $\pm$ 0.0124 dex/kpc for the youngest ages. The [Al/Fe] gradient starts around $\approx$$-$0.0500 dex/kpc before flattening for more recent ages with very slightly positive gradients for stars younger than 3.66 Gyr. The [K/Fe] gradient has a U-shaped trend in its evolution, but its extremum point is between 5.58--8.36 Gyr ago, with these stars having a gradient of $-$0.0202 $\pm$ 0.0108 dex/kpc. 
The central abundance values of the [X/Fe] abundance ratios show that the central value remains mostly constant regardless of the radial slope (see Figure \ref{fig:smc_oddz_radxfe}).

For the odd-Z [X/H] gradients, both [Na/H] and [K/H] exhibit a U-shaped trend, while the [Al/H] flattens over time (see Figure \ref{fig:smc_all_grads_h}). The [Na/H] has a potential extremum point at the same time as most other elements while the [K/Fe] extremum point happens for stars aged 5.58--8.36 Gyr.


The [X/Mg] gradients for the odd-Z elements evolve differently (see Figure \ref{fig:smc_all_grads_mg}). Much like the [Na/Fe] and [Na/H], the [Na/Mg] gradient starts steep and positive and becomes shallower over time. Sometime between 2.23 and 3.66 Gyr ago the [Na/Mg] gradient inverts. The [Al/Mg] gradient shows a U-shaped trend in its evolution with a potential extremum point for the stars in the 3.66--5.58 Gyr age bin. Its trend is reminiscent of the shape seen for [Na/H]. The [K/Mg] gradient evolutionary trend shows an overall downward tendency, but oscillates as it flattens.


  

\subsubsection{Iron Peak Elements}
\label{ssec:smc_ironpeak_evolve}

Multiple [X/Fe] gradients show a potential U-shaped trend except for [V/Fe] (see Figure \ref{fig:smc_all_grads_fe}). While the trend is not as prominent, it is visible for [Cr/Fe], [Mn/Fe], and [Ni/Fe]. Both [Cr/Fe] and [Mn/Fe] have an extremum point around the 3.66--5.58 Gyr age bin, while the heavier iron peak elements ([Co/Fe] and [Ni/Fe]) have an extremum point earlier around the 5.58--8.36 Gyr age bin. The radial trends show that the central abundances for the lighter iron peak elements are fairly constant over time (see Figure \ref{fig:smc_ironpeak_radxfe}). This is less true for the heavier iron peak elements and [Fe/H]. 


Many of the iron peak elements do exhibit a U-shaped trend with an extremum point at the same time as most of the other elements that also have a U-shaped trend (see Figure \ref{fig:smc_all_grads_h}). This is true for [Cr/H], [Mn/H], [Fe/H], and [Ni/H] as well as potentially [Co/H]. The [V/H] gradient's evolution shows a lot of large, sporadic excursions and, therefore, it is challenging to say what the gradient truly is doing over time. 



The [X/Mg] gradients for Cr, Mn, Co, and Ni have similar evolutionary trends, but V and Fe are outliers (see Figure \ref{fig:smc_all_grads_mg}). The [Cr/Mg], [Mn/Mg], [Co/Mg], and [Ni/Mg] all show a downward trend.
All of these start positive for the oldest stars, but eventually they invert and become negative. The [V/Mg] gradient shows an evolution similar to the [V/H] gradient evolution. The [Fe/Mg] does have a critical point in its evolution for stars aged 3.66--5.58 Gyr.

\subsubsection{The Neutron Capture Element Cerium}
\label{ssec:smc_neutron_evolve}

For the [Ce/Fe] gradient, the oldest two age bins have very negative values compared to the other age bins (see Figure \ref{fig:smc_all_grads_fe}). The radial trends show the dispersion in the [Ce/Fe] tends to be on the larger side (see Figure \ref{fig:smc_sr_radxfe}). The flattening and then subsequent steepening in the gradient evolution can also be seen in the radial trends quite clearly. This contrasts with what is seen in Figure \ref{fig:smc_all_grads_h}, where the gradient potentially shows the U-shaped trend with an extremum point in the 3.66--5.58 Gyr age bin. The [Ce/Mg] gradient does show temporal evolution, though it is fairly complex. 





\subsection{Age-[X/Fe] Trends}
\label{ssec:smc_age_xfe_trends}

Overall, the age-[X/Fe] trends for the SMC are mostly flat, though some show an increase in abundance for younger ages. For each of the plots pertinent to this section (Figures \ref{fig:smc_cn_axfe}, \ref{fig:smc_ind_alpha_axfe}, \ref{fig:smc_comb_alpha_axfe}, \ref{fig:smc_oddz_axfe}, \ref{fig:smc_ironpeak_axfe}, and \ref{fig:smc_sr_axfe}), if an age-[X/Fe] is above the black line then that region is overabundant and if the trend is below then the converse is true. The black line is the fiducial age-[X/Fe] trend and was found without any spatial binning unlike the colored lines. 





\subsubsection{Carbon \& Nitrogen}
\label{sssec:smc_cn_age_xfe_trends}

\begin{figure*}
    \centering
    \includegraphics[width=0.85\textwidth]{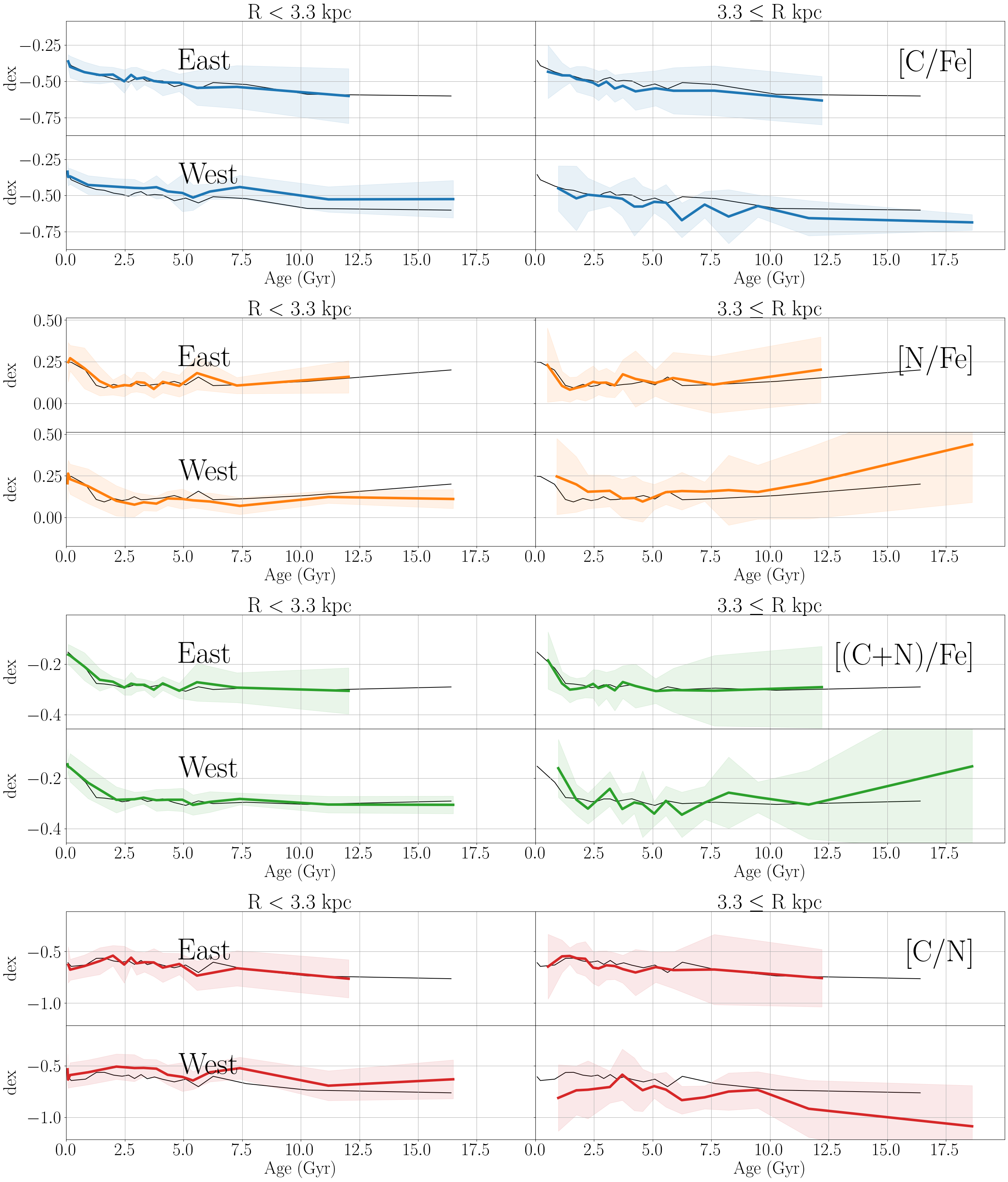}
    \caption{The age-[X/Fe] trends for the SMC for carbon and nitrogen. The top row in each panel shows the age-[X/Fe] trends for the northern part of the SMC and the bottom row shows the trends in the south. The SMC is split into two radial bins that contain an equal number of stars. For each sub-panel the age-[X/Fe] trend is included with a band of 1$\sigma$ in addition the black fiducial line shows the age-[X/Fe] trend without any binning. Clearly the [N/Fe] and [(C+N)/Fe] have a sharp increase for recent times.}
    \label{fig:smc_cn_axfe}
\end{figure*}

The age-[X/Fe] trends for the carbon and nitrogen abundances are mostly flat (see Figure \ref{fig:smc_cn_axfe}). The [C/Fe] abundance in the SMC very gradually increases over time throughout the galaxy. The western side of the galaxy for radii $<$ 3.3 kpc is overabundant with the trend falling above the fiducial line. The outer galaxy is underabundant in both the east and west. 

The age-[N/Fe] trend reveals that throughout the SMC there was an increase in [N/Fe] within the last Gyr or so. Also, the outer galaxy is nitrogen-rich in both the eastern and western side for old stars, though the west is more nitrogen-rich. The inner west tends to be close to the fiducial or slightly underabundant. The [(C+N)/Fe] age-[X/Fe] trend looks analogous to the [N/Fe] trend because it matches the fiducial throughout the galaxy and the outer west is the most different.



An interesting feature with the age-[C/N] trend is the underabundance for all ages in the west for radii $>$3.3 kpc, but the overabundance for the inner radii of the west. There may also be a very slight hump between 0.0--5.0 Gyr. 





\subsubsection{\texorpdfstring{$\alpha$}\,-Elements}
\label{sssec:smc_alpha_age_xfe_trends}

Overall, the individual $\alpha$-elements have very flat age-[X/Fe] trends with somewhat larger deviations from this behaviour at recent times (see Figure \ref{fig:smc_ind_alpha_axfe}). The lightest individual $\alpha$-element is [O/Fe] and the age-[O/Fe] trends in the east of the SMC follow the global fiducial, for the most part. There is actually a slight hump in the fiducial trendline and this is seen in the inner eastern region as well. The west is more complex with the younger stars in the inner region being oxygen-poor. The outer western region shows that the oldest stars are much more overabundant in [O/Fe] and the youngest stars are also slightly overabundant, but the ages between $\sim$5.5 Gyr and $\gtrsim$ 7.5 Gyr are underabundant in oxygen.

The age-[Mg/Fe] trend shows that the stars with radii $<$ 3.3 kpc have a downturn trend in magnesium. The outer SMC in the east has a flat age-[Mg/Fe] trend. The outer west has an age-[Mg/Fe] trend that is similar to the age-[O/Fe] trend in the same age bin, but the magnitude of the difference in the trends are smaller than what is seen in [O/Fe].

The next lightest individual $\alpha$-element is silicon. Both the inner and outer regions have flat age-[X/Fe] trends in the east. The inner western SMC appears to be deficient in silicon for ages younger than 5 Gyr.  The [Si/Fe] in the outer west looks very similar to the age-[Mg/Fe] trend in the same area.

Looking at the [S/Fe] and [Ca/Fe] age-abundance trends, both are very flat and do not show much deviation from the fiducial trendline or bin-to-bin. The [Ti/Fe] abundance is also quite flat like these abundances, except there is a slight tendency to increase over time.

The composite $\alpha$-abundance age-[X/Fe] trends look similar to the individual abundance ones (see Figure \ref{fig:smc_comb_alpha_axfe}). The age-[$\alpha$/Fe] trend shows that the eastern SMC follows the fiducial trendline better than in the west. There is also a bump present in the abundance for ages less than 5 Gyr. For younger stars, with ages that are $\lesssim$2.5 Gyr in the inner western SMC, there is a deficiency in $\alpha$-abundance. Stars with similar ages in the outer west show an overabundance of [$\alpha$/Fe], while between $\sim$5 Gyr to $\sim$ 12.0 Gyr the stars are [$\alpha$/Fe]-poor. However, the oldest stars are [$\alpha$/Fe]-rich.

The age-[$\alpha_\text{h}$/Fe] trend in the east, at all radii, looks a lot like the age-[$\alpha$/Fe] trend. This is also the case for stars at radii $<$ 3.3 kpc in the west as the age-[$\alpha_\text{h}$/Fe] trend looks like the analogous trend in [$\alpha$/Fe]. The outer west of the SMC has a [$\alpha_\text{h}$/Fe] trend similar to the outer west for [$\alpha$/Fe], but the dip in [$\alpha_\text{h}$/Fe] for intermediate ages only happens $\sim$4 Gyr to just past 7.5 Gyr. The explosive $\alpha$-element abundance, [$\alpha_\text{ex}$/Fe], has similar behaviour to [$\alpha_\text{h}$/Fe], but the age-[X/Fe] trend is flatter. 

The age-[$\alpha_\text{h}$/$\alpha_\text{ex}$] trend shows a slight downturn in the inner galaxy for radii $<$ 3.3 kpc with a deficiency for younger stars in the west. The outer east has a slight overabundance compared to the fiducial trendline. Finally, the outer west SMC [$\alpha_\text{h}$/$\alpha_\text{ex}$] looks like [$\alpha_\text{h}$/Fe], but with more exaggerated deviations from the fiducial trendline. 



\subsubsection{Odd-Z Elements}
\label{sssec:smc_oddz_age_xfe_trends}

For the odd-Z elements, the eastern SMC has essentially flat age-[X/Fe] trends for [Na/Fe], [Al//Fe], and [K/Fe] (see Figure \ref{fig:smc_oddz_axfe}). This is also true for stars with radii $<$ 3.3 kpc for [Al/Fe] and [K/Fe]. This not the case with the inner SMC which is [Na/Fe]-poor especially for ages over 5 Gyr. The outer galaxy age-[Na/Fe] trend shows that it is rich in [Na/Fe] for most of the lifetime of the SMC. The western region for radii larger than 3.3 kpc shows that the galaxy is poor in [Al/Fe] for ages between $\sim$4 Gyr to $\sim$8 Gyr. The outer western part of the SMC has a more complicated age-[K/Fe] trend, but the average trend appears to somewhat follow the fiducial trendline.

\subsubsection{Iron Peak Elements}
\label{sssec:smc_iron_age_xfe_trends}

Much like other groups of elements, the iron peak age-[X/Fe] trends are also mostly flat, though some deviate from this pattern (see Figure \ref{fig:smc_ironpeak_axfe}). The [V/Fe] abundance for the youngest stars in the inner galaxy show an increasing abundance with age, while this is less so for the outer galaxy. The eastern outer galaxy does show this increase more than in the west, but nowhere near the inner galaxy. In the west, for stars of the inner galaxy that are older than $\sim$12.5 Gyr, show a deficiency in [V/Fe], but the outer western galaxy shows an overabundance for most ages.

The age-[Cr/Fe] trend stays close to solar for all times and radii $<$3.3 kpc. This is also the case for the stars at larger radii on the eastern side. The outer SMC in the west shows a different behaviour with a [Cr/Fe] deficiency for ages $\sim$10 Gyr.  The age-[Mn/Fe] trend looks similar to the [Cr/Fe] one. The average value is less then solar and is around $-$0.2 dex. The outer west is [Cr/Fe] poor for ages $\lesssim$ 8 Gyr.

There is a steady increase from $\approx$ $-$1.5 dex up to $\approx$ $-$0.6 at the current day for the inner galaxy. The outer eastern region also shows this trend although it only reaches up to $\approx$ $-$1.0 dex at the present time. The outer west is the most divergent from the rest of the galaxy as evidenced by the deviation from the fiducial trendline. While the outer west does show an increase in [Fe/H], around 2.5 Gyr the trend turns down showing a decrease in [Fe/H] for that part of the galaxy. Also, the outer west galaxy is metal-poor compared to the fiducial trendline for the older stars. 

The age-[Co/Fe] trend is very flat staying slightly under the solar value. This is also case for [Ni/Fe], which is just under solar as well and follows the fiducial. The greatest deviations from the fiducial trendline happen in the outer west of the SMC. The average of the trends still follow the fiducial for both [Co/Fe] and [Ni/Fe]. 




\subsubsection{The Neutron Capture Element Cerium}
\label{sssec:smc_neutron_age_xfe_trends}

Looking at the age-[Ce/Fe] trend in Figure \ref{fig:smc_sr_axfe}, it is clear that the SMC [Ce/Fe] abundance increases over time with a more substantial increase for ages less than 5 Gyr. The eastern part of the SMC is neither deficient nor overabundant in cerium as there is almost no difference between the global fiducial trend and those for each radial bin. On the other hand, the west shows an overabundance in cerium for the inner SMC for ages $\gtrsim$ 5 Gyr, but the outer galaxy is underabundant.





\section{Discussion}
\label{sec:smc_discussion}

Below, we discuss the implications of our APOGEE abundance gradient results for the SMC. First is the overall metallicity gradient of the SMC and how it compares to literature values, second what the evolution of the calculated gradients and age-abundance trends, and finally how the SMC compares to the LMC and MW.




\subsection{SMC Metallicity Gradient}
\label{ssec:smc_feh_grad_discuss}

Here the metallicity ([Fe/H]) gradient calculated with APOGEE is compared to many other literature sources. Most of the literature sources gradients in units of dex/deg and to compare to this work, these have been converted to units of dex/kpc\footnote{We multiply by $\frac{180}{62.44\pi}$ deg/kpc and make use of the small angle approximation.}. Also, the literature sources discussed here are not exhaustive
but demonstrate the range of measured values in the SMC radial metallicity gradient.

\begin{table}
	\centering
	\caption{A table of only the [Fe/H] gradients calculated for the SMC in this work to make it easier to refer back to the previously calculated values.}
	\label{tab:smc_feh_recall}
	\begin{tabular}{ccc} 
		\hline
		Age Bin & $\nabla_\text{R}$ & Error \\
            (Gyr) & (dex/kpc) & (dex/kpc) \\
		\hline
            overall & $-$0.0546 & 0.0043 \\
		t $\leq$ 2.23 & $-$0.0742 & 0.0068 \\
		2.23 $<$ t $\leq$ 3.66 & $-$0.0529 & 0.0057 \\
		3.66 $<$ t $\leq$ 5.58 & $-$0.0286 & 0.0076 \\
            5.58 $<$ t $\leq$ 8.36 & $-$0.0466 & 0.0092 \\
            8.36 $<$ t & $-$0.0335 & 0.0139 \\
		\hline
	\end{tabular}
\end{table}

\begin{figure}
    \centering
    \includegraphics[scale=0.3]{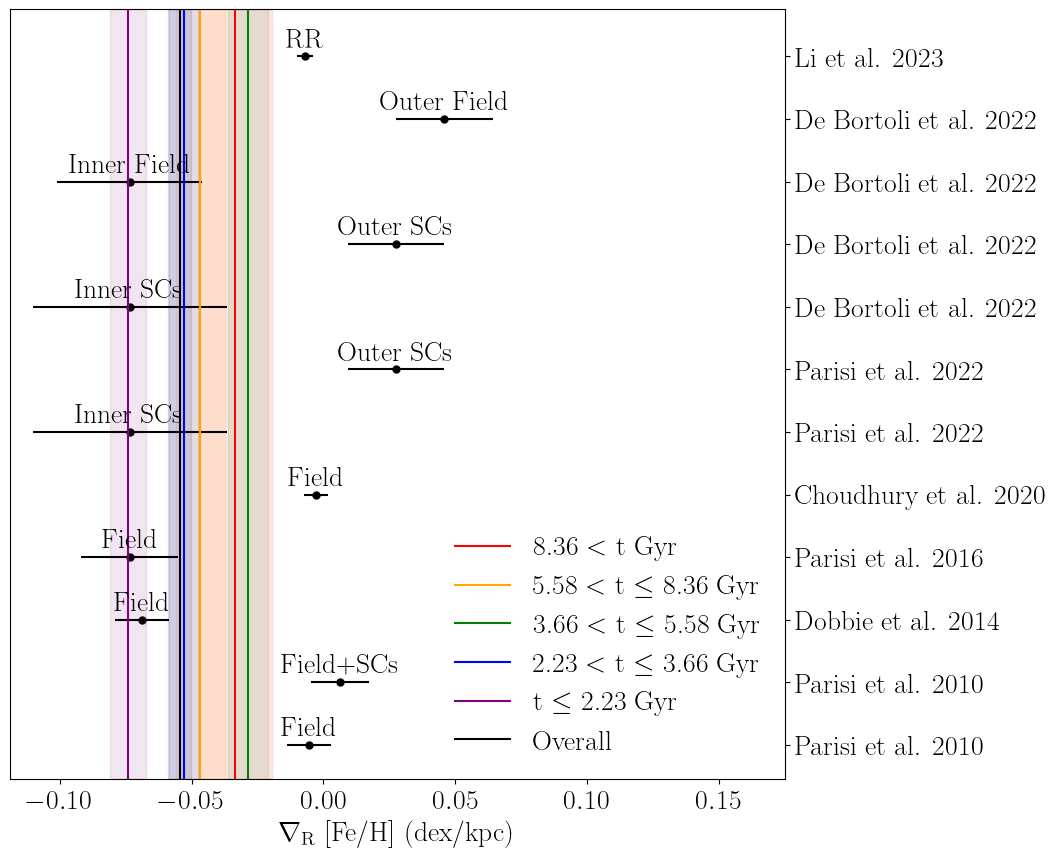}
    \caption{A comparison of the [Fe/H] gradients found in this work to multiple literature values. The vertical lines and bands correspond to the gradients for each of the age bins with the associated errors (see Table \ref{tab:smc_feh_recall}). The literature values are shown as points. The gradients calculated by \protect\cite{debortoli2022smcgrad} out to 3.4$^\circ$ overlap with the overall gradient as well as values found by \protect\cite{dobbie2014smcgrad} and \protect\cite{parisi2016smcgrad}. Around half of the literature gradients overlap with any of the gradients calculated in this work.}
    
    \label{fig:smc_lit_compare_feh}
\end{figure}

The overall metallicity gradient gradient of the SMC in this work was found to be $-$0.0546 $\pm$ 0.0043 dex/kpc, which overlaps with gradients found by \citet{dobbie2014smcgrad}, \citet{parisi2016smcgrad}, \citet{parisi2022cat} and \citet{debortoli2022smcgrad}. See Figure \ref{fig:smc_lit_compare_feh}, Tables \ref{tab:smc_lit_grads} for a quantitative comparison to these and other literature sources. For convenience Table \ref{tab:smc_feh_recall} contains every single [Fe/H] calculated for the different age bins.

\cite{parisi2010smcgrad} investigated the metallicity of RGB stars calculated from the near-infrared Ca {\scriptsize II} Triplet (CaT) for various samples out to a radius of $\sim$7.86$^\circ$. Looking at exclusively field stars, the gradient was found to be $-$0.0055 $\pm$ 0.0083 dex/kpc and this overlaps with the gradient of the oldest stars found in this work. Interestingly, if star clusters are included with the field stars, then the gradient moves away from the calculated gradients in this work becoming $+$0.0064 $\pm$ 0.0110 dex/kpc. 

Looking at the RGB stars out to 5$^\circ$ in the galaxy, \citet{dobbie2014smcgrad} find quite different results from \cite{parisi2010smcgrad}. Although \cite{dobbie2014smcgrad} also used field stars and CaT metallicities, they found a gradient of $-$0.0688 $\pm$ 0.0101 dex/kpc, which is consistent with the gradient of the youngest stars in the APOGEE sample.

Revisiting the gradient of the SMC, \cite{parisi2016smcgrad} find a value of $-$0.0734 $\pm$ 0.0184 dex/kpc using CaT-derived metallicities for field stars. This gradient does not overlap with the overall metallicity gradient found herein, but it does with the youngest metallicity gradient and also the \cite{dobbie2014smcgrad} value.

\citet{choudhury2020smcgrad} use a different metallicity-estimation technique  to explore the metallicity distribution of the SMC. Metallicities were derived by relating the slope of the RGB branch in the $Y$ vs.\ $Y - K_{\rm s}$ color-magnitude diagram (CMD). The slopes were calculated for 1,180 subregions of the SMC. The calculated gradient using the RGB slope metallicities is $-$0.0028 $\pm$ 0.0046 dex/kpc, which is similar to what was found by \cite{parisi2010smcgrad} for both the field stars and field stars+star clusters. 

Recently \cite{parisi2022cat} again revisited the metallicity gradient of the SMC using the CaT technique. The inner (R < 3.4$^\circ$) metallicity gradient was found to be $-$0.0734 $\pm$ 0.0367 dex/kpc, while the outer (R > 3.4$^\circ$) clusters have a gradient of $+$0.0275 $\pm$ 0.0184 dex/kpc. Clearly the inner cluster gradient overlaps with metallicity gradient found in this work.

The CaT technique is used to derive metallicities in \cite{debortoli2022smcgrad}. In that work, they explore the gradient of the inner (R < 3.4$^\circ$) and outer (R > 3.4$^\circ$) SMC using both field stars and star clusters. Both of the outer gradients fall far outside the range of any of the gradients calculated in this work. With the outer field stars having a gradient of $+$0.0459 $\pm$ 0.0184 dex/kpc and the outer clusters having a slightly shallower gradient of $+$0.0275 $\pm$ 0.0184 dex/kpc. Both of the inner gradients overlap with the overall metallicity gradient and many other ones calculated here. 

The final literature source discussed here is \cite{li2023photometric}. In that work, metallicities are found for RR Lyrae stars in Gaia by using $G$-metallicity relations and near--IR period--absolute magnitude--metallicity relations. The gradient found was $-$0.0070 $\pm$ 0.0030 dex/kpc, which is consistent with \citet{choudhury2020smcgrad}, \citet{parisi2010smcgrad}, and the old RGB star gradient calculated in this work.

This variation in the literature values for the SMC gradient is reminiscent of what is seen for the LMC gradient (Paper II). Interestingly, if there is agreement with the literature values it tends to be with the youngest gradients with potential overlap with the overall gradient, especially when considering the inner SMC gradients in literature (see Figure \ref{fig:smc_lit_compare_feh}). This may be due to a selection bias as there are more stars toward to center of the SMC and the fact that the center tends to be slightly younger on average. More data will be needed to make sure that these results are robust and to fill in the gaps between the APOGEE fields. Even with this, the metallicity gradient found in this work is the largest gradient study that uses high resolution spectra of individual stars.





\begin{table}
	\centering
	\caption{Table of literature values for the [Fe/H]-gradient of the SMC. All but \protect\cite{li2023photometric} and \protect\cite{choudhury2020smcgrad} use CaT equivalent widths to the [Fe/H]. }
	\label{tab:smc_lit_grads}
	\begin{tabular}{cccc}
		\hline
		  Source & $\nabla_{\rm R}$ & Method \\
                   & (dex/kpc) & \\
		\hline
            Parisi et al. 2010 & $-$0.0055 $\pm$ 0.0083 & CaT Field \\
            ---''--- & $+$0.0064 $\pm$ 0.0110 & CaT Field+SCs \\
            Dobbie et al. 2014 & $-$0.0688 $\pm$ 0.0101 & CaT Field \\
            Parisi et al. 2016 & $-$0.0734 $\pm$ 0.0184 & CaT Field \\
            Choudhury et al. 2020 & $-$0.0028 $\pm$ 0.0046 & RGB Slope \\
            Parisi et al. 2022 & $-$0.0734 $\pm$ 0.0367 & CaT Inner SCs \\
            ---''--- & $+$0.0275 $\pm$ 0.0184 & CaT Outer SCs \\
            De Bortoli et al. 2022 & $-$0.0734 $\pm$ 0.0367 & CaT Inner SCs \\
            ---''--- & $+$0.0275 $\pm$ 0.0184 & CaT Outer SCs \\
            ---''--- & $-$0.0734 $\pm$ 0.0275 & CaT Inner Field \\
            ---''--- & $+$0.0459 $\pm$ 0.0184 & CaT Outer Field \\
            Li et al. 2023 & $-$0.0070 $\pm$ 0.0030 & Gaia G Band RR \\
		\hline
	\end{tabular}
\end{table}

\subsection{Evolution of Abundance Gradients of the SMC}


Abundance gradients for 24 abundance ratios were calculated using 2,062 field RGB stars in the SMC. Combining these gradients with ages calculated using the method in Paper II makes it possible to study the evolution of each of the gradients.

While there are extremum points in the trends of the gradients for most elements, many have one for stars aged 3.66--5.58 Gyr. Abundances with this critical point appearing in their evolution includes [C/Fe], [N/Fe], [Na/Fe], [Cr/Fe], [Mn/Fe], most of the [X/H] gradients, [O/Mg], [Ca/Mg], [Na/Mg], [Mn/Mg], and [Fe/Mg]. The fact that this feature is so ubiquitous in many elements from different element groups, suggests a major galaxy-wide event caused it. 

In Paper II, it was found that extremum points in the evolution of LMC abundance gradients are possibly connected to starbursts from intergalactic interactions. It is known that an LMC-SMC interaction produced a starburst in each galaxy \citep{dobbie2014smcgrad,nidever20lazy,hasselquist2021satellites}. The starburst in the LMC was found to happen about 2 Gyr ago in \cite{nidever20lazy} and is consistent with the change in the abundance trends in Paper II. The turnover in the gradient evolution for the SMC was found to happen earlier around 3.66--5.58 Gyr ago. This is consistent with the starburst in \cite{dobbie2014smcgrad} 5--6 Gyr ago. \cite{hasselquist2021satellites} found that the SMC did experience a burst of star formation 3--4 Gyr before the LMC. Qualitatively the results of that paper agree with this present work, but the time of the burst is somewhat different even with using the same data. In that work the time of the starburst was determined by fitting chemical evolution models to the data. Also, \cite{massana2022synchronized} found bursts of star formation at $\sim$3 Gyr and $\sim$2 Gyr in the SMC, while the LMC star formation ramps up at $\sim$3 Gyr ago but doesn't peak until $\sim$2 Gyr ago. As previously mentioned, \cite{rubele2018sfh} finds a peak in the SFR between 3.98$-$6.31 Gyr ago. This result best matches what is seen with the U-shaped trend in the abundance gradients because 5 Gyr falls in the 3.66--5.58 Gyr age bin.  

It is not entirely clear why inconsistencies exist between many of the photometrically determined SFHs and the star formation bursts deduced from the chemistry. Obviously this is not the case with \cite{rubele2018sfh}.  Nevertheless, the other results qualitatively support the SMC experiencing a star formation burst that precedes the one in the LMC.

Radial migration is a process whereby stars move to a different radius because additional angular momentum is imparted on them \citep[e.g.][]{sellwood2002radial}. Naturally this will change the observed gradients especially for older stars so the present day gradients may not be representative of the gradients when the stars formed. \cite{lu2022migration} and \cite{ratcliffe2023mwgrads} find that while radial migration has affected the gradients in the MW, past interactions produce a signature steepening of gradients when it would be expected to see a flattening. But \cite{schroyen2013simulations} showed that dwarf galaxy metallicity gradients are less affected by radial migration than larger galaxies. This could be used to argue that the SMC gradients have changed less due to radial migration. Regardless it is best to keep in mind radial migration when ruminating about the results in this work and to think of the evolutionary trends as more age-dependent trends.

Additionally, in Paper II it was found that many of the age-[X/Fe] trends have a turnover with an extremum point due to an LMC starburst which further helps pinpoint the time frame of the burst. This is not what is seen in the SMC. The trends for the SMC are essentially flat with a few showing some slight increase or decrease for extremely young ages. There is no feature in the trends at the time of the burst. Most trends stay close to the fiducial trendline regardless of the spatial bin. This suggests that the SMC has experienced a slower enrichment history throughout the galaxy. Any disparities with the global fiducial trend are mostly in the outer western region of the galaxy.

\subsection{Comparing the SMC to the LMC and MW}
\label{ssec:smc_smc_lmc_mw}

In addition to comparing the calculated [Fe/H] gradient to other literature values, we also compare the SMC abundance gradients to those of the LMC and MW using the same method outlined in Section \ref{sec:smc_gradmethod} and the same APOGEE data. This allows us to conduct an apples-to-apples comparison of abundance gradients across a factor of $\sim$500 in total galaxy mass (SMC: $2 \times 10^{9}$ \msune, \citealt{Stanimirovic2004}; LMC: $1.38 \times 10^{11}$ \msune, \citealt{Erkal2019}; MW: $1.28 \times 10^{12}$ \msune, \citealt{Watkins2019}) in a uniform manner.
Gradients for the LMC were previously calculated using the APOGEE data in Povick et al.\ (2023b, in prep.).  We determine our own MW radial abundance gradients here.

MW RGB stars were selected from the APOGEE \texttt{allStar} file\footnote{\url{https://www.sdss.org/dr17/irspec/spectro_data/}} using a slightly modified version of the selection from \cite{jonsson2020apogee}:

\begin{enumerate}
    \item 1.5 $<$ $\log{g}$ $<$ 3.5
    \item {[C/N]} $>$ 0.04 - 0.46{[M/H]} - 0.0028$\cdot dT$
\end{enumerate}

\noindent
where

\[dT = {\rm T_{eff,spec}} - (4400 + 552.6\cdot(\log \text{g}_\text{spec} - 2.5) - 324.6\cdot\text{[M/H]})\]

\noindent
and where $\text{T}_\text{eff,spec}$, and $\log \text{g}_\text{spec}$ are the \texttt{TEFF\_SPEC} and \texttt{LOGG\_SPEC} columns in the \texttt{allStar} summary file, respectively. After this initial selection, a second selection was used to pick out stars close to the MW midplane using:

\begin{enumerate}
    \item 4.0 $\leq$ $R$ $\leq$ 15.0 kpc
    \item |$Z$| $\leq$ 1.0 kpc
\end{enumerate}

\noindent
where $R$ is the galactic radius of a star, and $Z$ is the vertical distance from the midplane. The selection criterion on $R$ was chosen so that there is large radial coverage, but that structures of the inner MW (i.e., bar and bulge) did not ``contaminate'' the disk sample. Once the selection is done, the method in Section \ref{sec:smc_gradmethod} is used to calculate the radial abundance gradients. Each of the galaxy gradients are scaled by the exponential disk radial scale length of each galaxy to remove the bias of different galaxy sizes. The radial scale of the disk for the SMC is 1.25 kpc \citep{gonidakis2009structure}, for the LMC is 1.667 kpc \citep{choi2018smashing}, and for the MW is 2.15 kpc \citep{bovy2013direct}.

It important to mention that the APOGEE LMC and SMC samples should be relatively free from MW contamination in that the number of MW stars is negligible. The contamination was minimized using selection cuts in ASPCAP stellar parameters, radial velocities, and proper motions from Gaia DR2 \citep{gaia2018dr2} as outlined in \cite{nidever20lazy}. In that paper it is stated that if a star in the MC sample is actually from the MC, then most likely it will be a halo star. Since the MW halo is made up of older metal poor stars this would skew the metallicity in the MCs to be more metal poor depending on how substantial the fraction of MW stars is.

\begin{figure*}
    \centering
    \includegraphics[width=\textwidth]{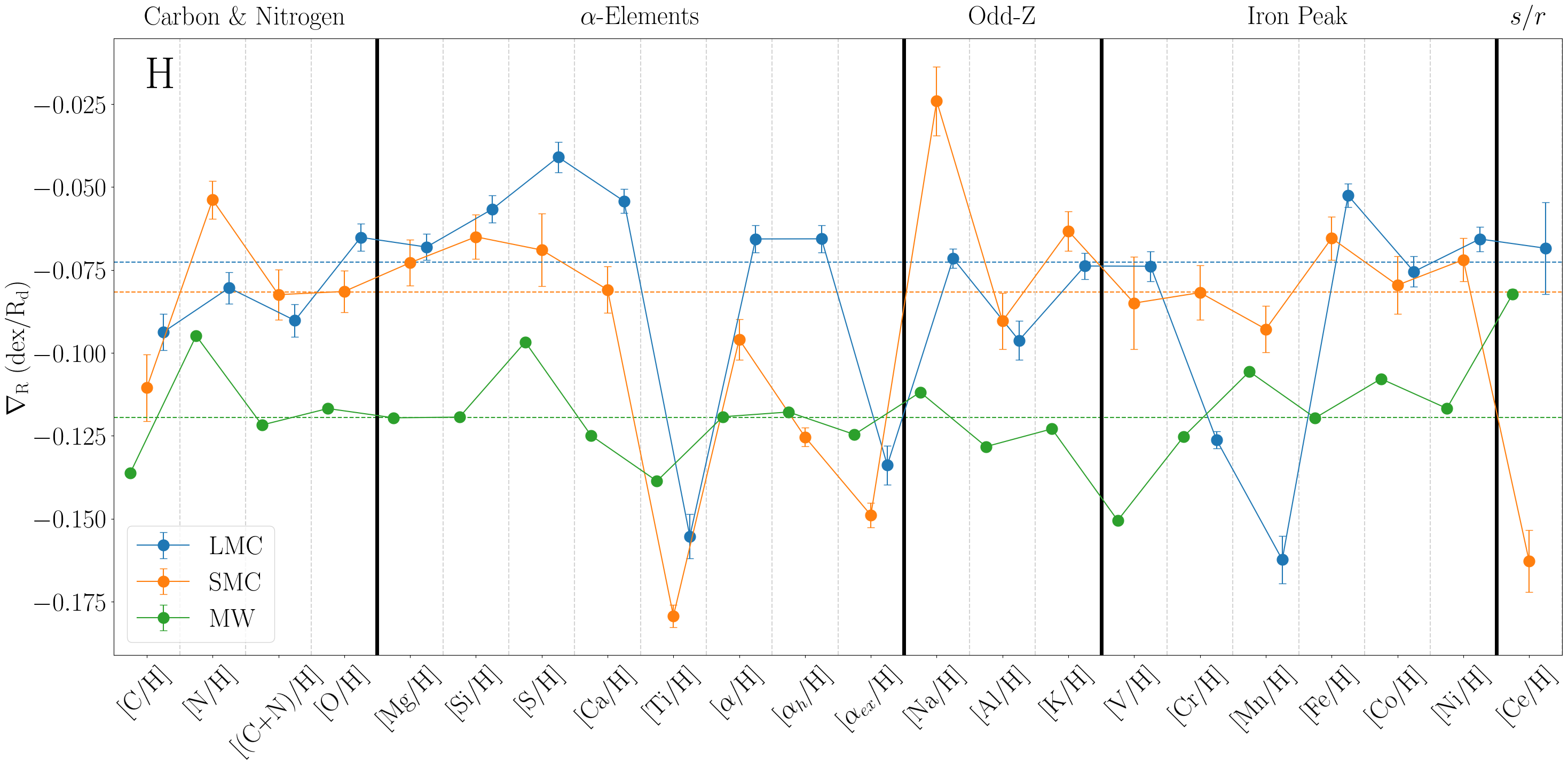}    
    \includegraphics[width=\textwidth]{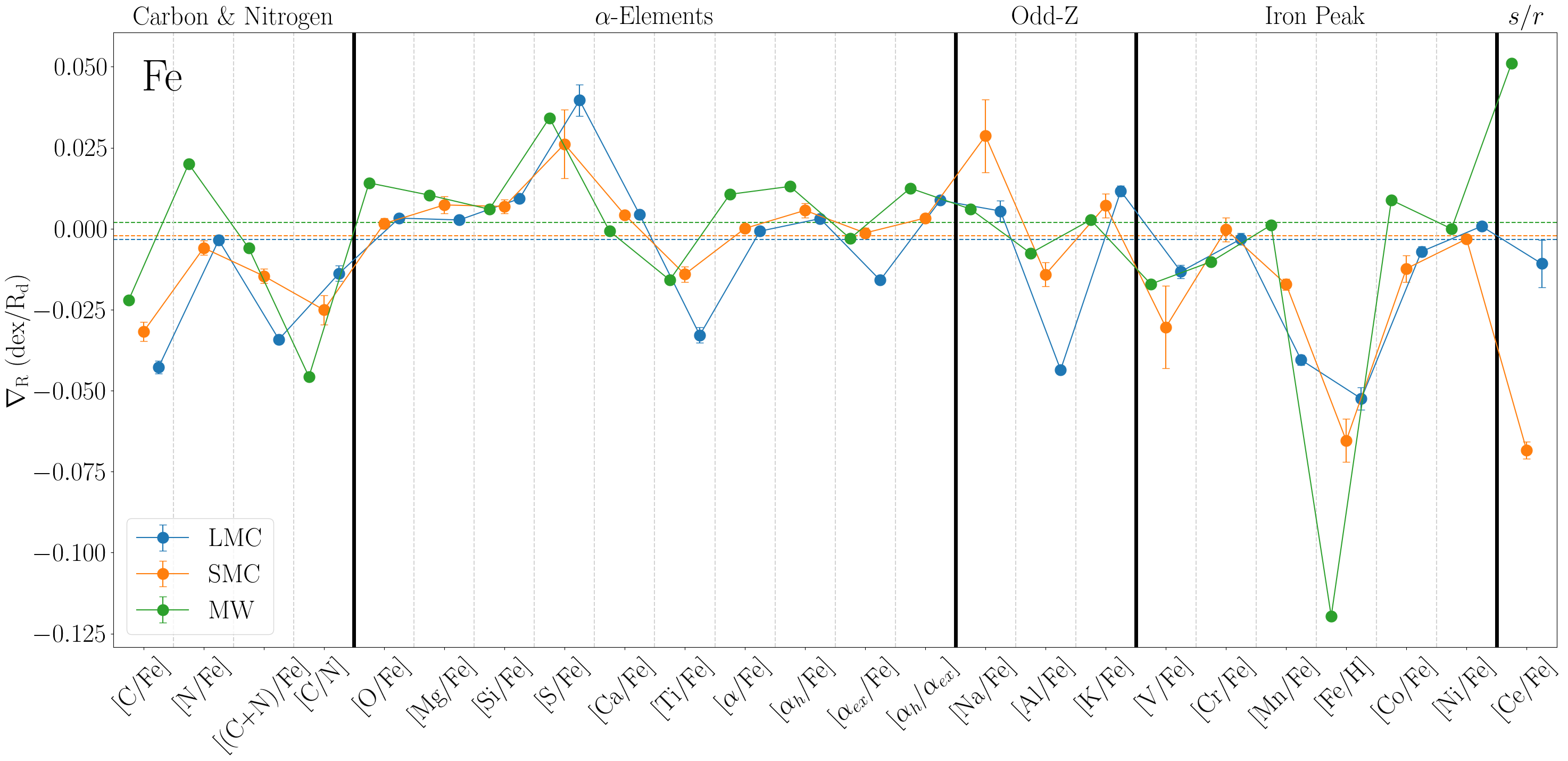}
    \caption{A comparison of the overall radial abundance gradients for the SMC, LMC and the MW. The LMC is shown in blue, the SMC in orange, and the MW in green. ({\em Top}) [X/H] gradients which show a clear separation between the MW and the dwarf galaxies.  The MW has median abundance gradients (dotted horizontal lines) almost 50\% larger then ($-$1.20 dex/$R_d$) than the MCs ($-$0.80 dex/$R_d$).  ({\em Bottom}) [X/Fe] gradients where all three galaxies show much more consistent patterns and almost no difference in the median gradients.}
    \label{fig:smc_smc_lmc_mw}
\end{figure*}

In the top panel of Figure \ref{fig:smc_smc_lmc_mw}, the SMC, LMC, and MW really stand out from each other in their [X/H] gradients. The LMC gradients appear the shallowest followed by the SMC and then the MW. While not exactly the same, both of the Clouds have more similar gradients between themselves versus the MW. The clear separation between the Clouds and the MW for the [X/H] suggest that there may exist a mass dependence in these gradients. The median [X/Mg] gradients for the Clouds are $\sim -$0.075 dex/R$_\text{d}$ and the MW median gradient is roughly 50\% steeper with a value of $-$0.120 dex/R$_\text{d}$. 

In contrast, the patterns in the [X/Fe] gradients of the three galaxies look very similar (bottom panel of Figure \ref{fig:smc_smc_lmc_mw}). There are only a few elements for which a substantial difference occurs such as the case with [Ce/Fe]. For some elements, the gradients for the Clouds look alike (e.g. [N/Fe], [C/N], [O/Fe], [$\alpha$/Fe], etc.) while there are elements for which the LMC and MW look alike or the SMC and the MW do. Since the Clouds and the MW have had a very different accretion and formation history, the overall consistency of the patterns in the radial [X/Fe] gradients (in dex/$R_\text{d}$) may suggest that these are a fairly universal aspect of galaxies.

As for the [X/Mg] gradients (Figure \ref{fig:smc_smc_lmc_mw_mg}), the three galaxies have comparable gradients except in a few elements. The largest separation between the galaxies occurs with [Fe/Mg] and [Ce/Mg] and is similar in many ways to what is seen for the [X/H] gradients. Possibly these gradients are galaxy mass sensitive. The median [X/Mg] gradients for all three galaxies is flat.



\section{Summary}
\label{sec:smc_summary}


We calculated radial abundance gradients for 24 abundance ratios and their temporal evolution with an MCMC technique using 2,062 APOGEE DR17 RGB stars in the SMC. In addition, age-abundance trends for [C/N] and [$\alpha_\text{h}/\alpha_\text{ex}$] were measured.


The SMC has an overall metallicity gradient of $-$0.0546 $\pm$ 0.0043 dex/kpc, which is steeper than the LMC, but nowhere near as steep as the MW field star gradient. Accounting for disk scale length, the metallicity gradient of the SMC is $-$0.0654 $\pm$ 0.0067 dex/R$_\text{d}$, the LMC is $-$0.0525 $\pm$ 0.0035 dex/R$_\text{d}$, and the MW is $-$0.1196 $\pm$ 0.0009 dex/R$_\text{d}$. The calculated SMC gradient agrees with gradients from \citet{dobbie2014smcgrad}, \citet{parisi2016smcgrad}, \citet{parisi2022cat} and \citet{debortoli2022smcgrad}. There is better agreement on the inner SMC metallicity gradient calculated in the literature, regardless of the source using GCs or field stars.

Further comparing the overall SMC gradients to the LMC and MW show that in many cases the gradients are too similar to distinguish between the three galaxies. This is true for the [X/Fe] and [X/Mg] gradients, but not with the [X/H] gradients. The radial iron gradients, regardless of the fiducial element, deviate from each other for all three galaxies, though the MCs have closer gradients. The median gradient for the [X/Fe] and [X/Mg] are very similar for all three galaxies, but this is not the case for [X/H]. For the [X/H] abundances, the median MW gradient is around $-$0.125 dex/R$_\text{d}$, while the Clouds have median gradients around $-$0.075 dex/R$_\text{d}$.



There are a substantial number of gradients with extremum points in their evolution for stars aged 3.66--5.58 Gyr old. Similar to what the LMC gradients in Paper II show, the extremum point in their evolution corresponds to the starburst triggered by a close interaction between the LMC and SMC, though the starburst in the SMC happens before the LMC. 


For the most part, the age-[X/Fe] trends are flat for the SMC, even more so than what was found in Paper II with the LMC. In most cases with any discernible differences, the outer SMC in the west seems to be more chemically different compared to the rest of the galaxy. Unfortunately, there is no feature in the trends which can help pinpoint the starburst due to the LMC-SMC interaction as was found in Paper II. 





\section*{Acknowledgements}


J.T.P. acknowledges support for this research from the National Science Foundation (AST-1908331) and the Montana Space Grant Consortium Graduate Fellowship.

D.L.N. also acknowledges support for this research from the National Science Foundation (AST-1908331).

S.R.M. and A.A. acknowledge support from the National Science Foundation grant AST-1909497.

D.G. gratefully acknowledges the support provided by Fondecyt regular n. 1220264.
D.G. also acknowledges financial support from the Direcci\'{o}n de Investigaci\'{o}n y Desarrollo de
la Universidad de La Serena through the Programa de Incentivo a la Investigaci\'{o}n de
Acad\'{e}micos (PIA-DIDULS).

R. R. M. gratefully acknowledges support by the ANID BASAL project FB210003.

Funding for the Sloan Digital Sky Survey IV has been provided by the Alfred P. Sloan Foundation, the U.S. Department of Energy Office of Science, and the Participating Institutions. 

SDSS-IV acknowledges support and resources from the Center for High Performance Computing  at the University of Utah. The SDSS website is www.sdss.org.

SDSS-IV is managed by the Astrophysical Research Consortium for the Participating Institutions of the SDSS Collaboration including the Brazilian Participation Group, the Carnegie Institution for Science, Carnegie Mellon University, Center for Astrophysics | Harvard \& Smithsonian, the Chilean Participation Group, the French Participation Group, Instituto de Astrof\'isica de Canarias, The Johns Hopkins University, Kavli Institute for the Physics and Mathematics of the Universe (IPMU) / University of Tokyo, the Korean Participation Group, Lawrence Berkeley National Laboratory, Leibniz Institut f\"ur Astrophysik Potsdam (AIP), Max-Planck-Institut f\"ur Astronomie (MPIA Heidelberg), Max-Planck-Institut f\"ur Astrophysik (MPA Garching), Max-Planck-Institut f\"ur Extraterrestrische Physik (MPE), National Astronomical Observatories of China, New Mexico State University, New York University, University of Notre Dame, Observat\'ario Nacional / MCTI, The Ohio State University, Pennsylvania State University, Shanghai Astronomical Observatory, United Kingdom Participation Group, Universidad Nacional Aut\'onoma de M\'exico, University of Arizona, University of Colorado Boulder, University of Oxford, University of Portsmouth, University of Utah, University of Virginia, University of Washington, University of Wisconsin, Vanderbilt University, and Yale University.

This work has made use of data from the European Space Agency (ESA) mission
{\it Gaia} (\url{https://www.cosmos.esa.int/gaia}), processed by the {\it Gaia}
Data Processing and Analysis Consortium (DPAC,
\url{https://www.cosmos.esa.int/web/gaia/dpac/consortium}). Funding for the DPAC
has been provided by national institutions, in particular the institutions
participating in the {\it Gaia} Multilateral Agreement.

\textit{Software:} Astropy \citep{pricewhelan2018astropy, robitaille2013astropy}, Matplotlib \citep{hunter2007matplotlib}, NumPy \citep{harris2020numpy}, SciPy \citep{virtanen2020scipy}

\section*{Data Availability}

All APOGEE DR17 data used in this study is publicly available and can be found at: \url{https://www.sdss4.org/dr17/data_access/}.




\bibliographystyle{mnras}
\bibliography{bibliography} 




\appendix

\section{Gaia ADQL Query}\label{app:smc_gaia_adql}

Here is the Gaia ADQL query used to get a sample of SMC stars from which the RC is selected for determining the APOGEE field RC proper motions. \\ 

\noindent
\texttt{SELECT * FROM gaiadr3.gaia\_source\_lite as g \\
WHERE 1 = CONTAINS(POINT(g.ra,g.dec), \\ 
CIRCLE(13.18333333,$-$72.82833333,11)) \\
AND g.phot\_g\_mean\_mag$<$20.5 \\
AND g.parallax IS NOT NULL \\
AND g.phot\_bp\_mean\_mag IS NOT NULL \\
AND g.phot\_rp\_mean\_mag IS NOT NULL \\
AND g.phot\_g\_mean\_mag IS NOT NULL \\ 
AND g.parallax$<$0.2 \\
AND g.parallax$>$$-$0.2 \\
AND g.pmra$>$=$-$3.0 \\
AND pmra$<$=3.0 \\
AND g.pmdec$>$=$-$3.0 \\
AND pmdec$<$=3.0 \\
AND g.phot\_g\_mean\_flux\_over\_error$>$3.0 \\
AND g.phot\_bp\_mean\_flux\_over\_error$>$3.0 \\
AND g.phot\_rp\_mean\_flux\_over\_error$>$3.0} \\

\section{\texorpdfstring{$\nabla_R$}\,[X/H] \& \texorpdfstring{$\nabla_R$}\,[X/Mg] Tables}\label{app:smc_grads_h_mg}

The tables for the [X/H] and [X/Mg] gradients, which do not appear in the main body of the paper.

\begin{table*}
	\centering
	\caption{Table of the [X/H] gradients for each of the age bins. Horizontal lines have been added marking the division of the previously defined groups of elements. In general descending down the table corresponds to an increase in atomic number. Each age bin has its own separate column.}
	\label{tab:smc_evolve_grad_h}
	\begin{tabular}{cccccc}
        \hline
         & t $\leq$ 2.23 & 2.23 $<$ t $\leq$ 3.66 & 3.66 $<$ t $\leq$ 5.58 & 5.58 $<$ t $\leq$ 8.36 & 8.36 $<$ t \\
        Element & $\nabla_\text{R}$ & $\nabla_\text{R}$ & $\nabla_\text{R}$ & $\nabla_\text{R}$ & $\nabla_\text{R}$ \\
         & (dex/kpc) & (dex/kpc) & (dex/kpc) & (dex/kpc) & (dex/kpc) \\
        \hline 
        {[C/H]} & $-$0.0812 $\pm$ 0.0081 & $-$0.0910 $\pm$ 0.0098 & $-$0.0440 $\pm$ 0.0117 & $-$0.1131 $\pm$ 0.0215 & $-$0.0776 $\pm$ 0.0241 \\
        {[N/H]} & $-$0.0647 $\pm$ 0.0082 & $-$0.0560 $\pm$ 0.0064 & $-$0.0244 $\pm$ 0.0055 & $-$0.0225 $\pm$ 0.0088 & $-$0.0258 $\pm$ 0.0110 \\
        {[(C+N)/H]} & $-$0.0891 $\pm$ 0.0068 & $-$0.0698 $\pm$ 0.0089 & $-$0.0292 $\pm$ 0.0075 & $-$0.0772 $\pm$ 0.0177 & $-$0.0262 $\pm$ 0.0150 \\
        \hline
        {[O/H]} & $-$0.0547 $\pm$ 0.0072 & $-$0.0591 $\pm$ 0.0054 & $-$0.0279 $\pm$ 0.0073 & $-$0.0513 $\pm$ 0.0082 & $-$0.0608 $\pm$ 0.0154 \\
        {[Mg/H]} & $-$0.0149 $\pm$ 0.0074 & $-$0.0597 $\pm$ 0.0072 & $-$0.0419 $\pm$ 0.0059 & $-$0.1092 $\pm$ 0.0086 & $-$0.0693 $\pm$ 0.0161 \\
        {[Si/H]} & $-$0.0307 $\pm$ 0.0059 & $-$0.0420 $\pm$ 0.0071 & $-$0.0204 $\pm$ 0.0056 & $-$0.0539 $\pm$ 0.0092 & $-$0.0618 $\pm$ 0.0142 \\
        {[S/H]} & $-$0.0236 $\pm$ 0.0096 & $-$0.0651 $\pm$ 0.0195 & $-$0.0486 $\pm$ 0.0072 & $-$0.0535 $\pm$ 0.0251 & $-$0.1362 $\pm$ 0.0441 \\
        {[Ca/H]} & $-$0.0561 $\pm$ 0.0068 & $-$0.0631 $\pm$ 0.0078 & $-$0.0461 $\pm$ 0.0077 & $-$0.0693 $\pm$ 0.0117 & $-$0.0662 $\pm$ 0.0175 \\
        {[Ti/H]} & $-$0.0786 $\pm$ 0.0081 & $-$0.0699 $\pm$ 0.0076 & $-$0.0371 $\pm$ 0.0102 & $-$0.0661 $\pm$ 0.0140 & $-$0.1352 $\pm$ 0.0141 \\
        {[$\alpha$/H]} & $-$0.0407 $\pm$ 0.0067 & $-$0.0622 $\pm$ 0.006 & $-$0.0344 $\pm$ 0.0071 & $-$0.0724 $\pm$ 0.0084 & $-$0.0657 $\pm$ 0.0141 \\
        {[$\alpha_\text{h}$/H]} & $-$0.0306 $\pm$ 0.0067 & $-$0.0607 $\pm$ 0.0061 & $-$0.0321 $\pm$ 0.0074 & $-$0.0677 $\pm$ 0.0094 & $-$0.0649 $\pm$ 0.0153 \\
        {[$\alpha_\text{ex}$/H]} & $-$0.0582 $\pm$ 0.0068 & $-$0.0586 $\pm$ 0.0066 & $-$0.0326 $\pm$ 0.0081 & $-$0.0658 $\pm$ 0.0118 & $-$0.0636 $\pm$ 0.0153 \\
        \hline
        {[Na/H]} & $-$0.0642 $\pm$ 0.0151 & $-$0.0288 $\pm$ 0.0164 & $-$0.0421 $\pm$ 0.0086 & $-$0.0267 $\pm$ 0.0140 & $-$0.0592 $\pm$ 0.0181 \\
        {[Al/H]} & $-$0.0305 $\pm$ 0.0050 & $-$0.0758 $\pm$ 0.0086 & $-$0.0663 $\pm$ 0.0103 & $-$0.1421 $\pm$ 0.0133 & $-$0.1379 $\pm$ 0.0175 \\
        {[K/H]} & $-$0.0134 $\pm$ 0.0047 & $-$0.0412 $\pm$ 0.0057 & $-$0.0413 $\pm$ 0.0020 & $-$0.0978 $\pm$ 0.0108 & $-$0.0417 $\pm$ 0.0198 \\
        \hline
        {[V/H]} & $-$0.1411 $\pm$ 0.0166 & 0.0260 $\pm$ 0.0061 & $-$0.0705 $\pm$ 0.0171 & $-$0.1005 $\pm$ 0.0168 & $-$0.0033 $\pm$ 0.0566 \\
        {[Cr/H]} & $-$0.0916 $\pm$ 0.0126 & $-$0.0700 $\pm$ 0.0114 & $-$0.0437 $\pm$ 0.0101 & $-$0.0486 $\pm$ 0.0151 & $-$0.0684 $\pm$ 0.0242 \\
        {[Mn/H]} & $-$0.0918 $\pm$ 0.0074 & $-$0.0747 $\pm$ 0.0095 & $-$0.0440 $\pm$ 0.0102 & $-$0.0562 $\pm$ 0.0136 & $-$0.046 $\pm$ 0.0161 \\
        {[Fe/H]} & $-$0.0741 $\pm$ 0.0070 & $-$0.0528 $\pm$ 0.0057 & $-$0.0286 $\pm$ 0.0076 & $-$0.0464 $\pm$ 0.0091 & $-$0.0336 $\pm$ 0.0139 \\
        {[Co/H]} & $-$0.0582 $\pm$ 0.0105 & $-$0.0631 $\pm$ 0.0072 & $-$0.0528 $\pm$ 0.0100 & $-$0.0675 $\pm$ 0.0201 & $-$0.0857 $\pm$ 0.0213 \\
        {[Ni/H]} & $-$0.0452 $\pm$ 0.0063 & $-$0.0597 $\pm$ 0.0068 & $-$0.0233 $\pm$ 0.0083 & $-$0.0575 $\pm$ 0.0107 & $-$0.0441 $\pm$ 0.0193 \\
        \hline
        {[Ce/H]} & $-$0.0963 $\pm$ 0.012 & $-$0.0587 $\pm$ 0.0107 & $-$0.0115 $\pm$ 0.0129 & $-$0.1380 $\pm$ 0.031 & $-$0.0884 $\pm$ 0.0298 \\
    \end{tabular}
\end{table*}

\begin{table*}
	\centering
	\caption{Table of the [X/Mg] gradients for each of the age bins. Horizontal lines have been added marking the division of the previously defined groups of elements. In general descending down the table corresponds to an increase in atomic number. Each age bin has its own separate column.}
	\label{tab:smc_evolve_grad_mg}
	\begin{tabular}{cccccc}
        \hline
         & t $\leq$ 2.23 & 2.23 $<$ t $\leq$ 3.66 & 3.66 $<$ t $\leq$ 5.58 & 5.58 $<$ t $\leq$ 8.36 & 8.36 $<$ t \\
        Element & $\nabla_\text{R}$ & $\nabla_\text{R}$ & $\nabla_\text{R}$ & $\nabla_\text{R}$ & $\nabla_\text{R}$ \\
         & (dex/kpc) & (dex/kpc) & (dex/kpc) & (dex/kpc) & (dex/kpc) \\
        \hline 
        {[C/Mg]} & $-$0.0500 $\pm$ 0.0054 & $-$0.0252 $\pm$ 0.0041 & $-$0.0221 $\pm$ 0.0062 & $-$0.0127 $\pm$ 0.0143 & $-$0.0093 $\pm$ 0.0072 \\
        {[N/Mg]} & $-$0.0742 $\pm$ 0.0058 & 0.0054 $\pm$ 0.0041 & 0.0239 $\pm$ 0.0037 & 0.0522 $\pm$ 0.0074 & 0.0285 $\pm$ 0.0036 \\
        {[(C+N)/Mg]} & $-$0.0556 $\pm$ 0.004 & $-$0.0132 $\pm$ 0.0027 & $-$0.0006 $\pm$ 0.0033 & 0.0215 $\pm$ 0.0070 & 0.0214 $\pm$ 0.0058 \\
        \hline
        {[O/Mg]} & $-$0.0196 $\pm$ 0.0035 & 0.0013 $\pm$ 0.0014 & 0.0012 $\pm$ 0.0008 & 0.0113 $\pm$ 0.0024 & 0.0111 $\pm$ 0.0024 \\
        {[Mg/H]} & $-$0.0154 $\pm$ 0.0072 & $-$0.0596 $\pm$ 0.0073 & $-$0.0418 $\pm$ 0.006 & $-$0.1094 $\pm$ 0.0086 & $-$0.0691 $\pm$ 0.0162 \\
        {[Si/Mg]} & $-$0.0106 $\pm$ 0.0032 & $-$0.0041 $\pm$ 0.0011 & 0.0035 $\pm$ 0.0018 & 0.0245 $\pm$ 0.0021 & 0.0167 $\pm$ 0.0027 \\
        {[S/Mg]} & 0.0097 $\pm$ 0.0109 & 0.0094 $\pm$ 0.0170 & $-$0.0349 $\pm$ 0.0111 & 0.0223 $\pm$ 0.0255 & $-$0.0262 $\pm$ 0.027 \\
        {[Ca/Mg]} & $-$0.0213 $\pm$ 0.0036 & $-$0.0057 $\pm$ 0.003 & 0.0153 $\pm$ 0.0024 & $-$0.0166 $\pm$ 0.0016 & 0.0108 $\pm$ 0.0066 \\
        {[Ti/Mg]} & $-$0.0429 $\pm$ 0.0042 & $-$0.0003 $\pm$ 0.0011 & $-$0.0003 $\pm$ 0.0053 & 0.0140 $\pm$ 0.0046 & $-$0.0545 $\pm$ 0.0031 \\
        {[$\alpha$/Mg]} & $-$0.0154 $\pm$ 0.0027 & $-$0.0025 $\pm$ 0.0012 & 0.0008 $\pm$ 0.0017 & 0.0111 $\pm$ 0.0034 & 0.0062 $\pm$ 0.0032 \\
        {[$\alpha_\text{h}$/Mg]} & $-$0.0097 $\pm$ 0.0018 & 0.0006 $\pm$ 0.0007 & 0.0006 $\pm$ 0.0004 & 0.0057 $\pm$ 0.0012 & 0.0055 $\pm$ 0.0011 \\
        {[$\alpha_\text{ex}$/Mg]} & $-$0.0214 $\pm$ 0.0033 & $-$0.0102 $\pm$ 0.0008 & 0.0000 $\pm$ 0.0025 & 0.0015 $\pm$ 0.0025 & 0.0043 $\pm$ 0.0030 \\
        \hline
        {[Na/Mg]} & $-$0.0302 $\pm$ 0.0160 & 0.0035 $\pm$ 0.0164 & 0.0121 $\pm$ 0.0109 & 0.0179 $\pm$ 0.0164 & 0.0603 $\pm$ 0.0234 \\
        {[Al/Mg]} & $-$0.028 $\pm$ 0.0042 & $-$0.0068 $\pm$ 0.0020 & $-$0.0133 $\pm$ 0.0041 & $-$0.0130 $\pm$ 0.0033 & $-$0.0283 $\pm$ 0.0069 \\
        {[K/Mg]} & $-$0.0085 $\pm$ 0.0043 & 0.0133 $\pm$ 0.0028 & $-$0.0008 $\pm$ 0.0028 & 0.0188 $\pm$ 0.0036 & 0.0088 $\pm$ 0.0011 \\
        \hline
        {[V/Mg]} & $-$0.1231 $\pm$ 0.0171 & $-$0.1002 $\pm$ 0.0119 & 0.0601 $\pm$ 0.0124 & $-$0.0737 $\pm$ 0.0221 & 0.0799 $\pm$ 0.0519 \\
        {[Cr/Mg]} & $-$0.0280 $\pm$ 0.0090 & $-$0.0066 $\pm$ 0.0078 & $-$0.0095 $\pm$ 0.0079 & $-$0.0017 $\pm$ 0.0153 & 0.0297 $\pm$ 0.0062 \\
        {[Mn/Mg]} & $-$0.0429 $\pm$ 0.0053 & $-$0.0254 $\pm$ 0.0044 & $-$0.0354 $\pm$ 0.0019 & 0.0036 $\pm$ 0.0055 & 0.0307 $\pm$ 0.0087 \\
        {[Fe/Mg]} & $-$0.0840 $\pm$ 0.0149 & $-$0.1194 $\pm$ 0.0138 & $-$0.0570 $\pm$ 0.0078 & $-$0.1530 $\pm$ 0.0185 & $-$0.1016 $\pm$ 0.0261 \\
        {[Co/Mg]} & $-$0.0177 $\pm$ 0.0055 & $-$0.0121 $\pm$ 0.0036 & $-$0.0126 $\pm$ 0.0082 & $-$0.0028 $\pm$ 0.0117 & 0.0070 $\pm$ 0.0093 \\
        {[Ni/Mg]} & $-$0.0098 $\pm$ 0.0030 & $-$0.0060 $\pm$ 0.0030 & 0.0111 $\pm$ 0.0039 & 0.0225 $\pm$ 0.0033 & 0.0215 $\pm$ 0.0055 \\
        \hline
        {[Ce/Mg]} & $-$0.0672 $\pm$ 0.0066 & $-$0.0065 $\pm$ 0.0076 & 0.0054 $\pm$ 0.009 & 0.0239 $\pm$ 0.0168 & $-$0.0364 $\pm$ 0.0044 \\
    \end{tabular}
\end{table*}

\section{Radius-Abundance-Age Trends}
\label{app:smc_age_xfe_radius_trends}

The radius-[x/Fe]-age trends for the $\alpha$-elements, odd-Z elements, iron peak elements and neutron capture elements that do not appear in the main body of the paper.

\begin{figure*}
    \centering
    \includegraphics[width=0.949\textwidth]{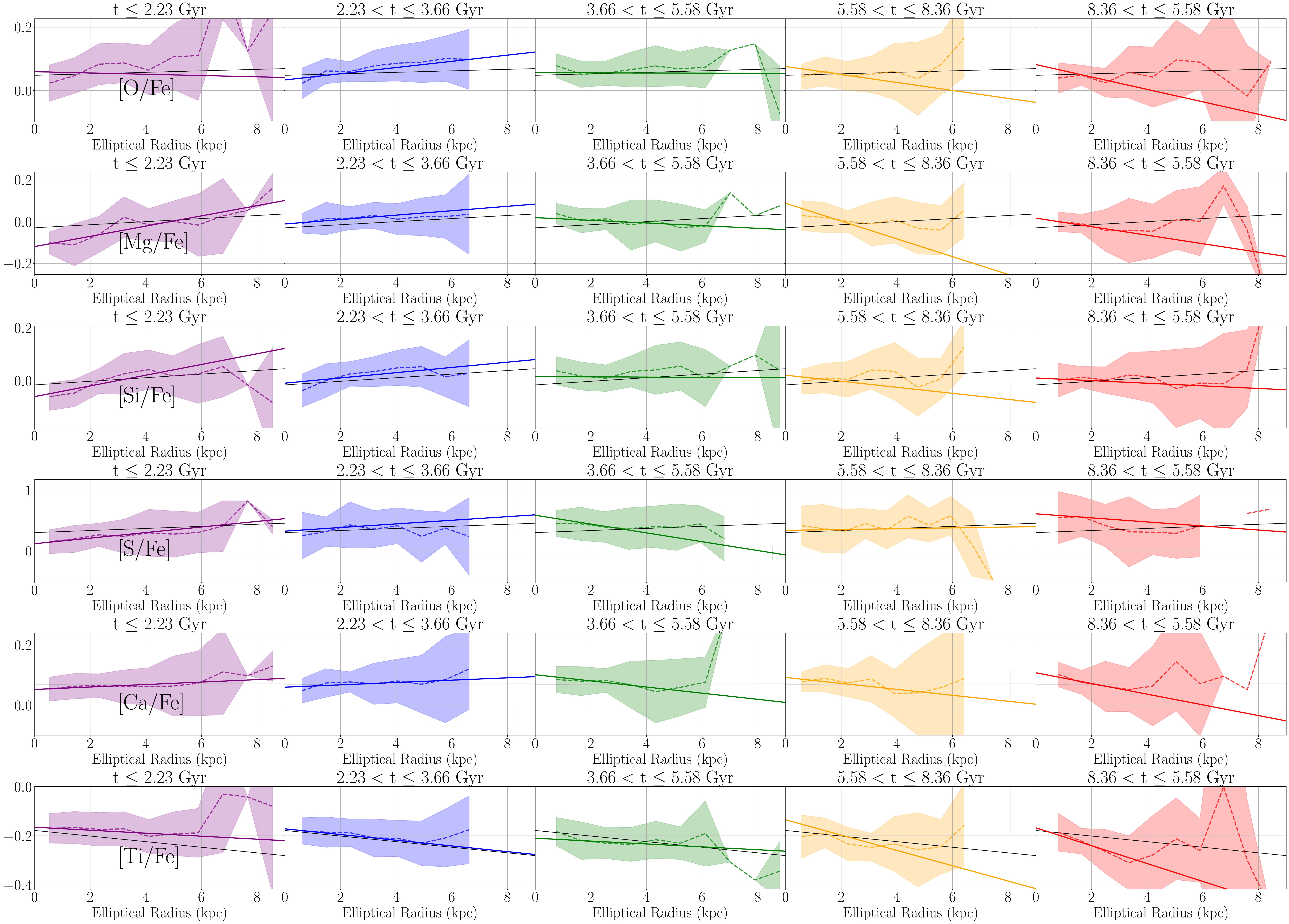}
    \caption{The same as for Figure~\ref{fig:smc_cn_radxfe}, but with the individual $\alpha$-element abundances.}
    \label{fig:smc_ind_alpha_radxfe}
\end{figure*}

\begin{figure*}
    \centering
    \includegraphics[width=0.95\textwidth]{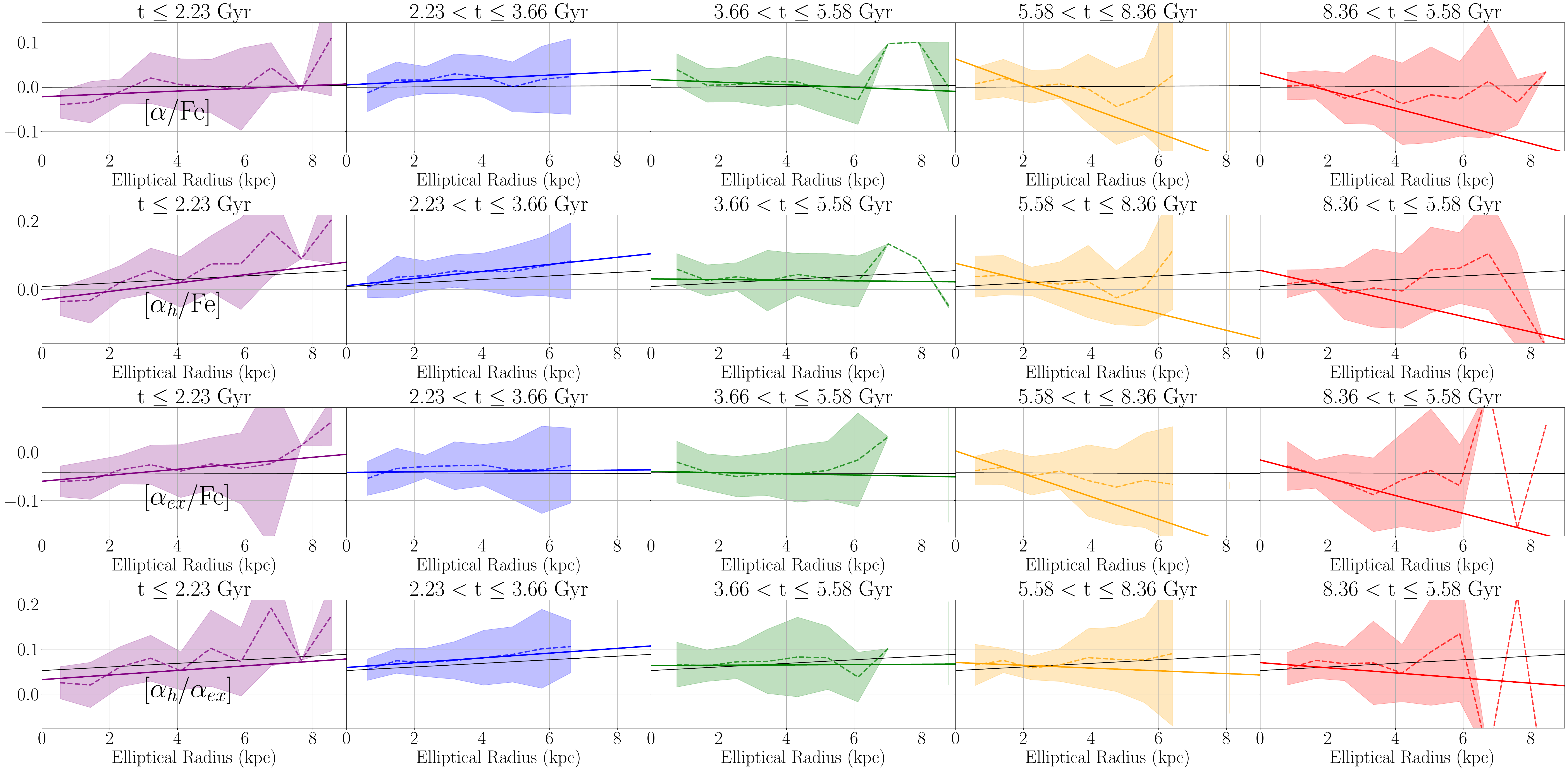}
    \caption{The radial abundance trends for the composite $\alpha$-element abundances.}
    \label{fig:smc_comb_alpha_radxfe}
\end{figure*}

\begin{figure*}
    \centering
    \includegraphics[width=0.95\textwidth]{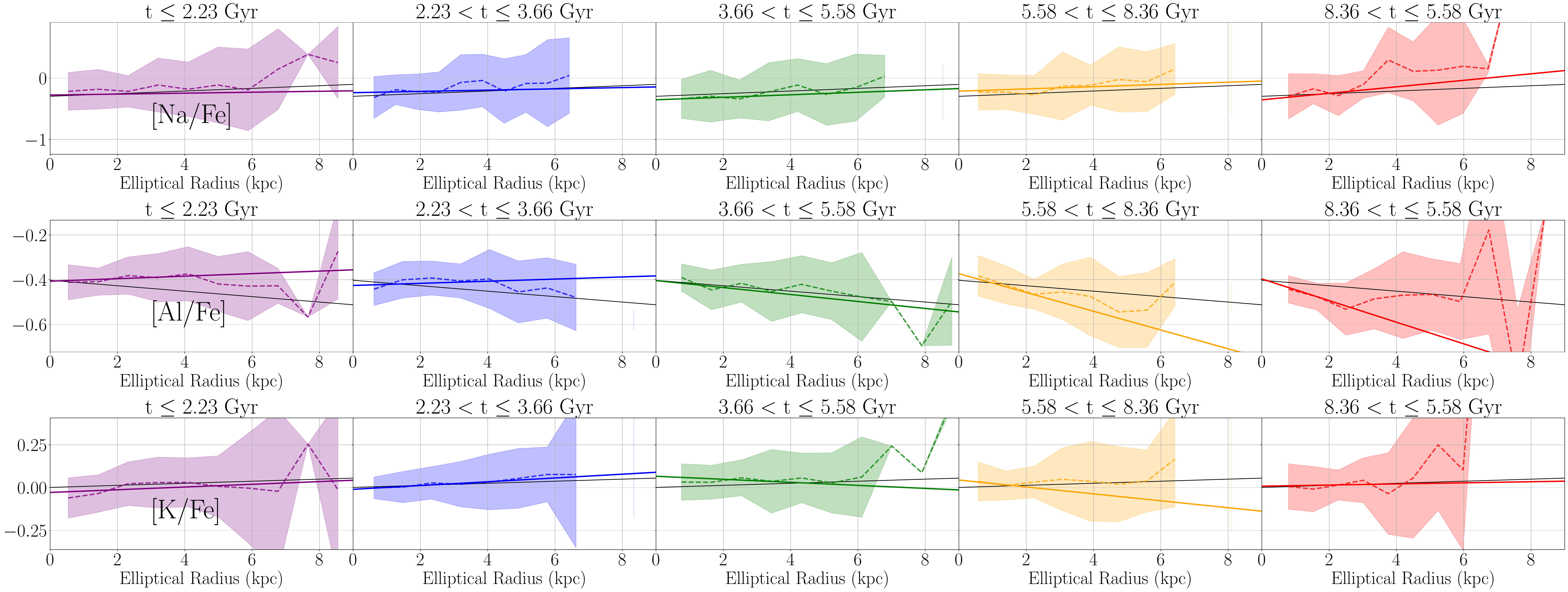}
    \caption{The radial abundance trends for the odd-Z abundances.}
    \label{fig:smc_oddz_radxfe}
\end{figure*}

\begin{figure*}
    \centering
    \includegraphics[width=0.95\textwidth]{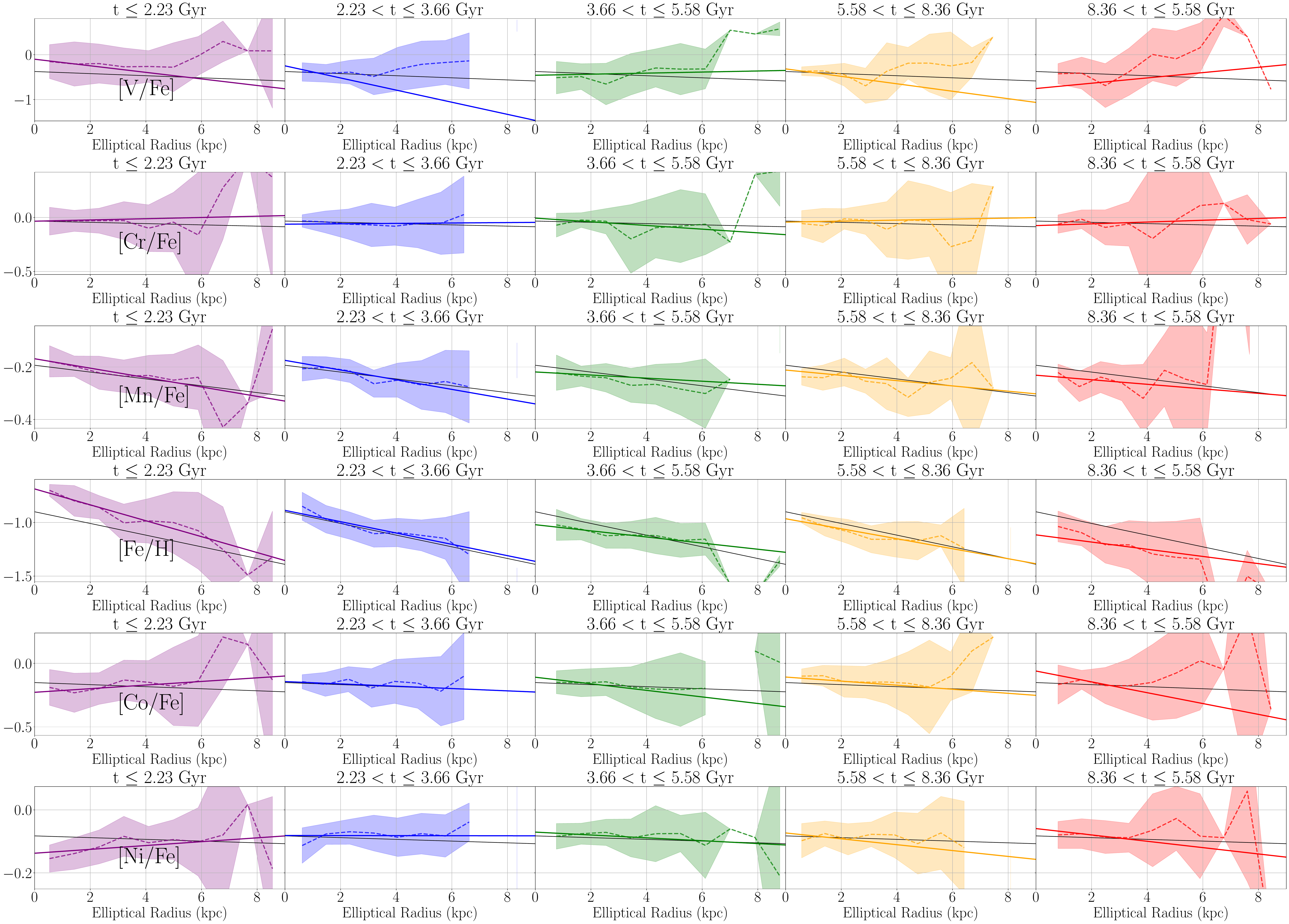}
    \caption{The radial abundance trends for the iron peak abundances.}
    \label{fig:smc_ironpeak_radxfe}
\end{figure*}

\begin{figure*}
    \centering
    \includegraphics[width=0.95\textwidth]{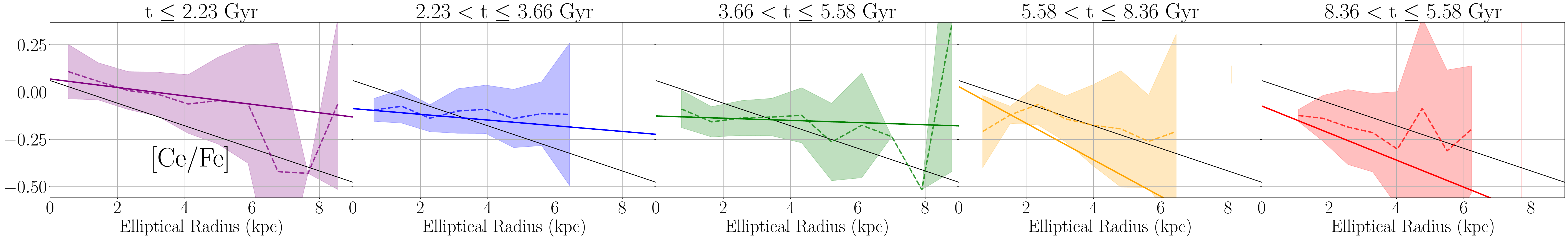}
    \caption{The radial abundance trends for the neutron capture abundances.}
    \label{fig:smc_sr_radxfe}
\end{figure*}

\section{age-[X/Fe] Trends}
\label{app:smc_age_xfe_trends}

The age-[X/Fe] trends for the $\alpha$-elements, odd-Z elements, iron peak elements and neutron capture elements that do not appear in the main body of the paper.

\begin{figure*}
    \centering
    \includegraphics[width=0.85\textwidth,height=0.9\textheight]{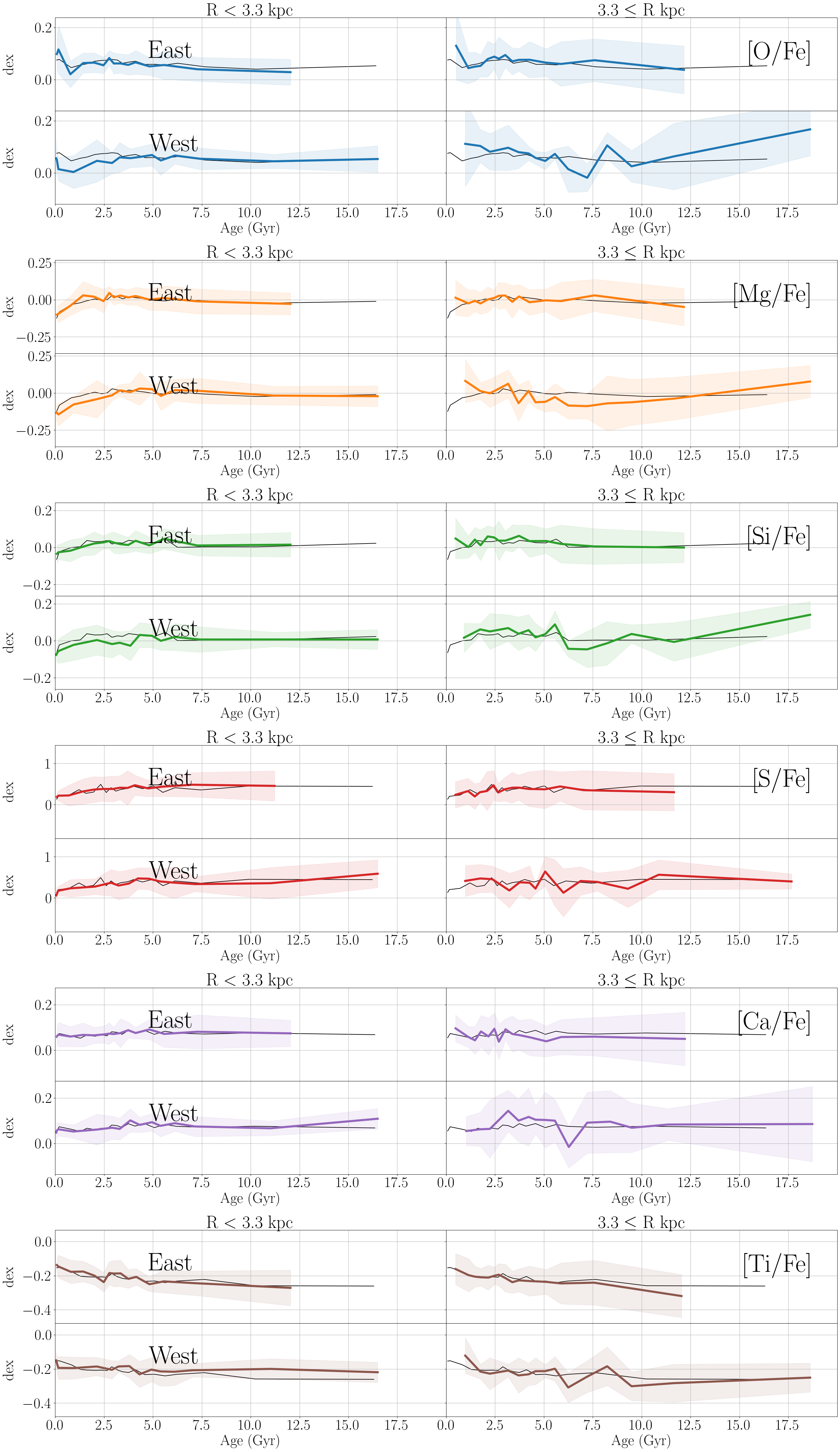}
    \caption{The age-[X/Fe] trends for the SMC for the individual $\alpha$-element abundances.}
    \label{fig:smc_ind_alpha_axfe}
\end{figure*}

\begin{figure*}
    \centering
    \includegraphics[width=0.875\textwidth]{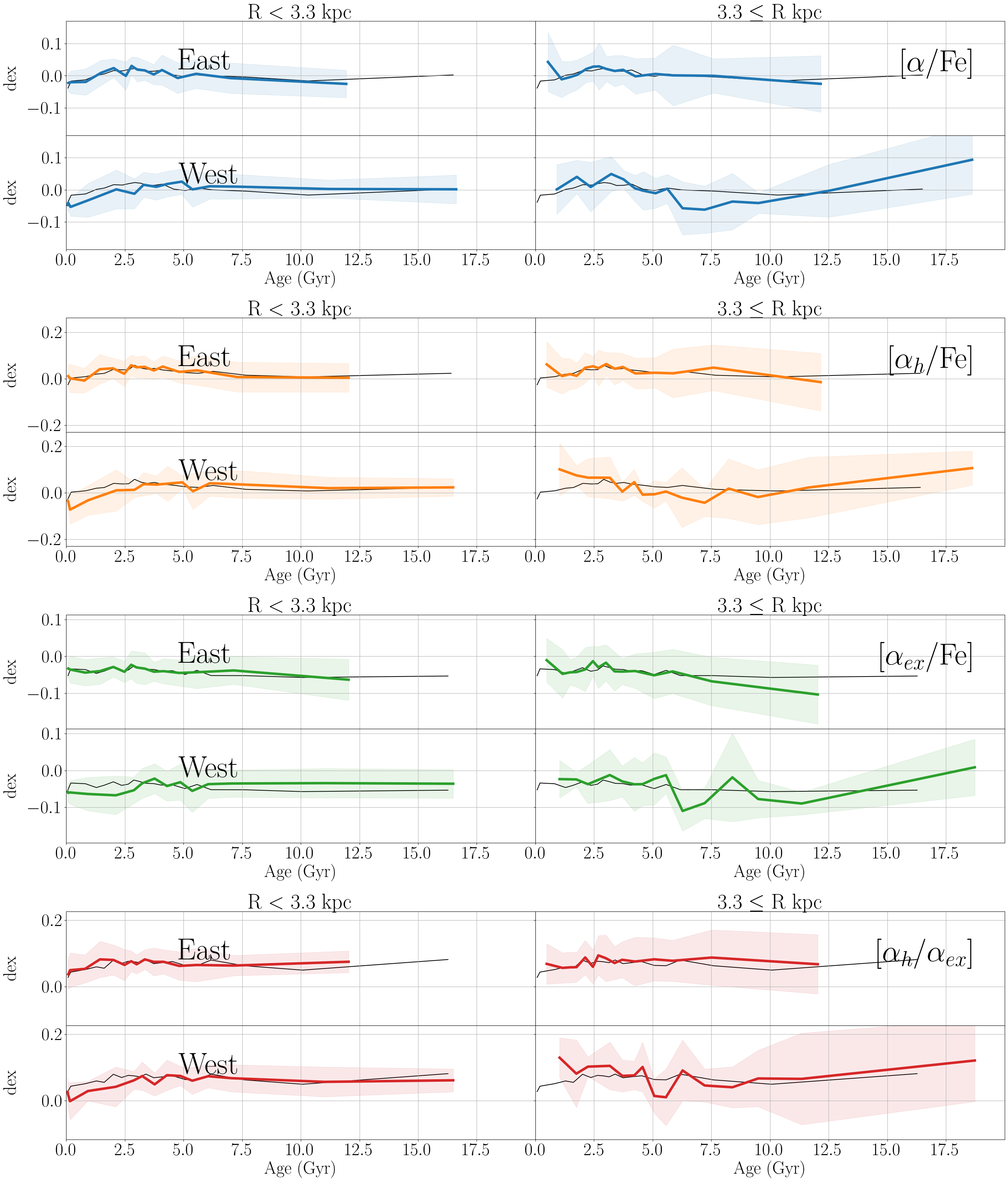}
    \caption{The age-[X/Fe] trends for the SMC for the composite $\alpha$-element abundances.}
    \label{fig:smc_comb_alpha_axfe}
\end{figure*}

\begin{figure*}
    \centering
    \includegraphics[width=0.95\textwidth]{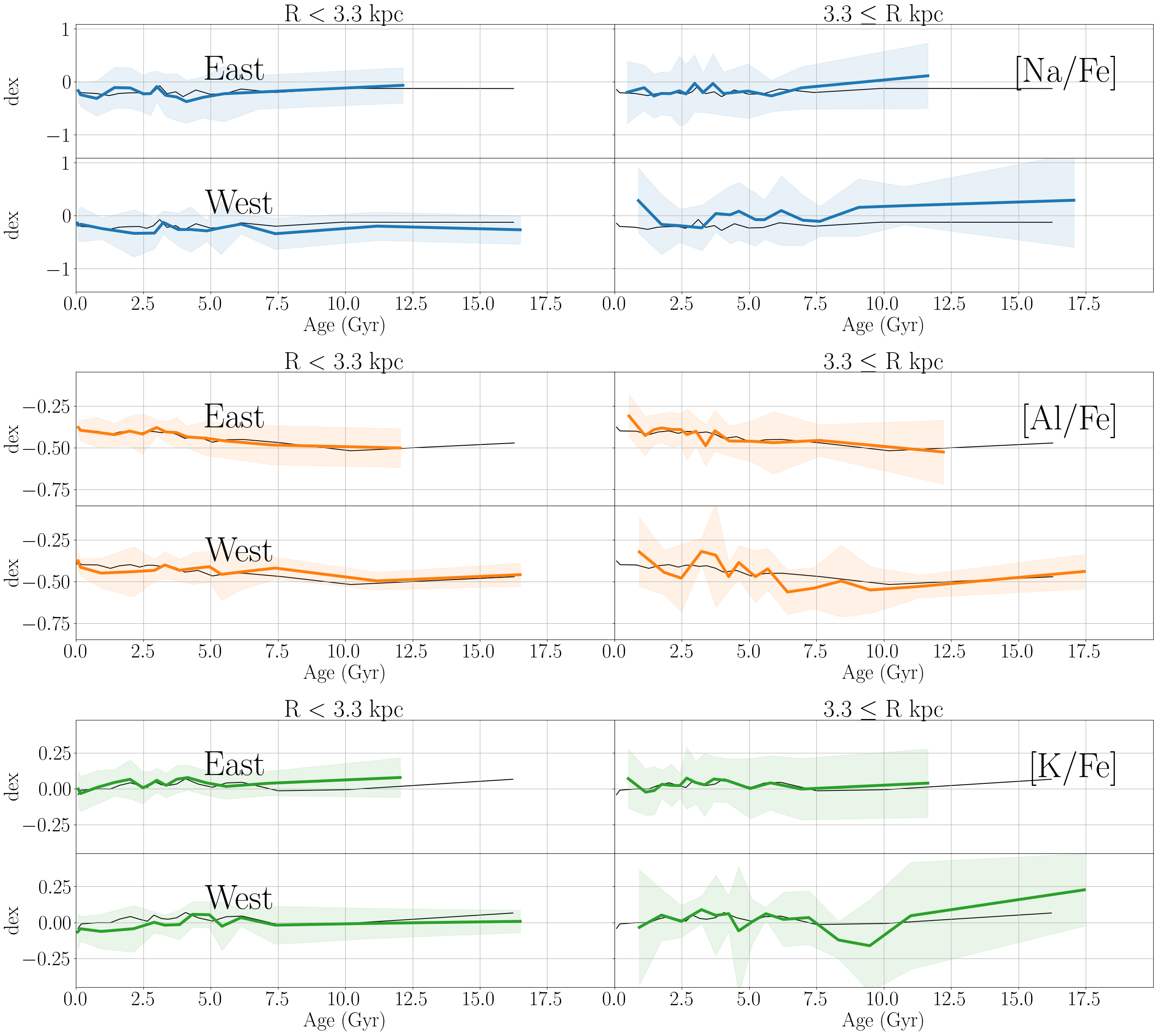}
    \caption{The age-[X/Fe] trends for the SMC for the odd-Z abundances.}
    \label{fig:smc_oddz_axfe}
\end{figure*}

\begin{figure*}
    \centering
    \includegraphics[width=0.9\textwidth,height=0.9\textheight]{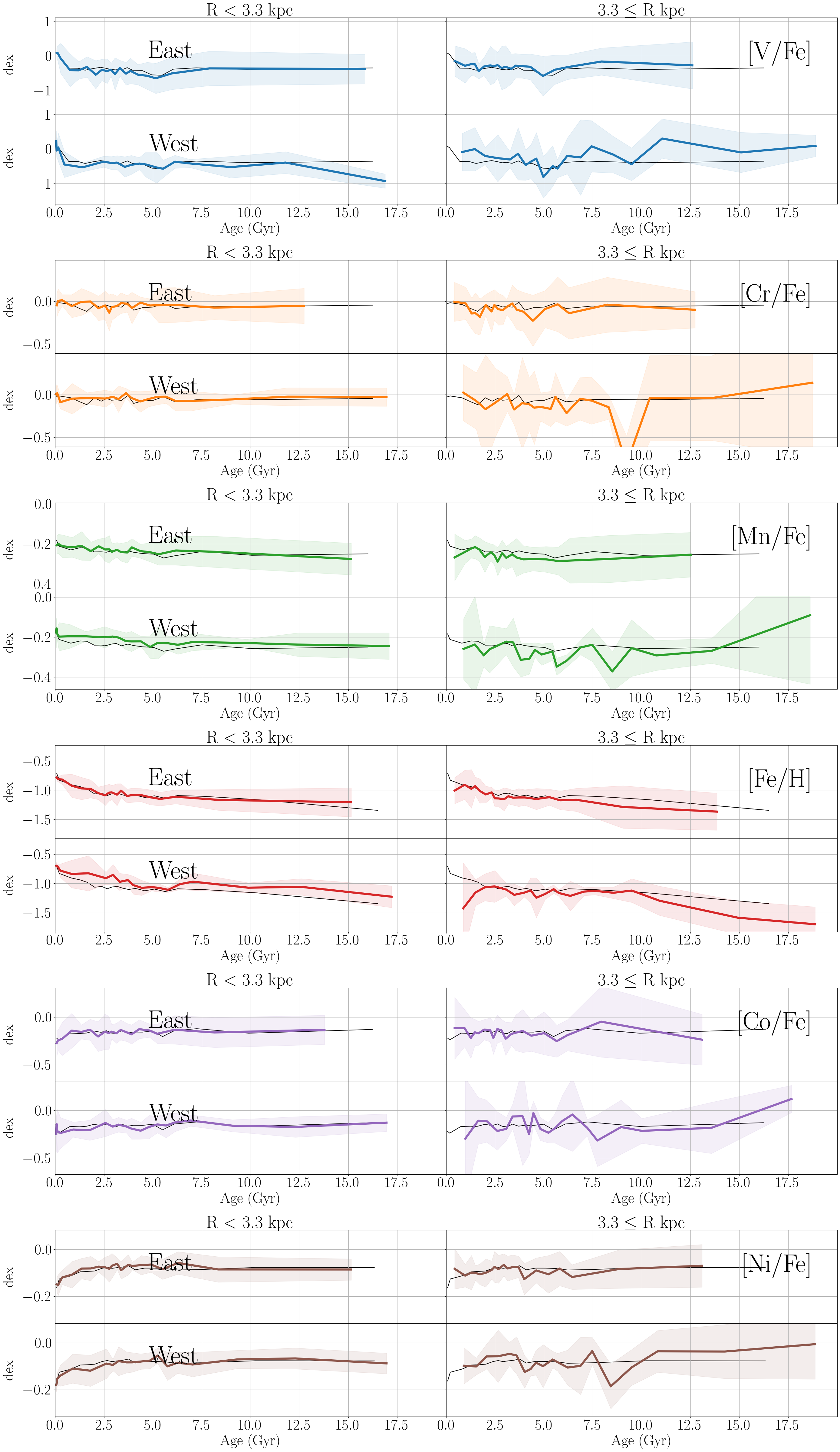}
    \caption{The age-[X/Fe] trends for the SMC for the individual iron peak abundances.}
    \label{fig:smc_ironpeak_axfe}
\end{figure*}

\begin{figure*}
    \centering
    \includegraphics[width=0.95\textwidth]{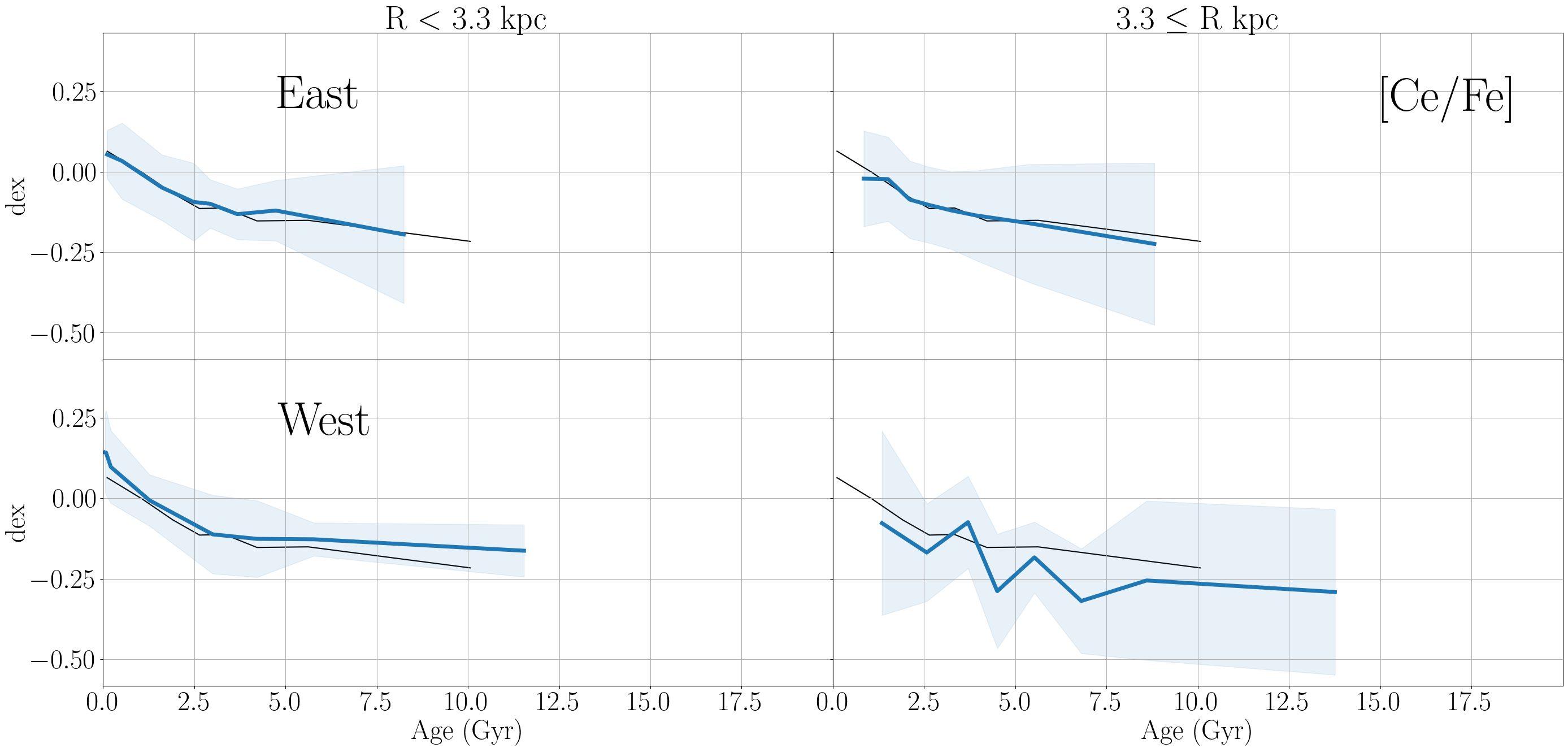}
    \caption{The age-[X/Fe] trend for the SMC for the neutron capture element cerium.}
    \label{fig:smc_sr_axfe}
\end{figure*}

\section{Comparing to the LMC and MW Extra Plots}\label{app:smc_smc_lmc_mw_extra}

Plots comparing the SMC, LMC, and Milky Way that do not appear in the main body of the paper.


\begin{figure*}
    \centering
    \includegraphics[width=\textwidth]{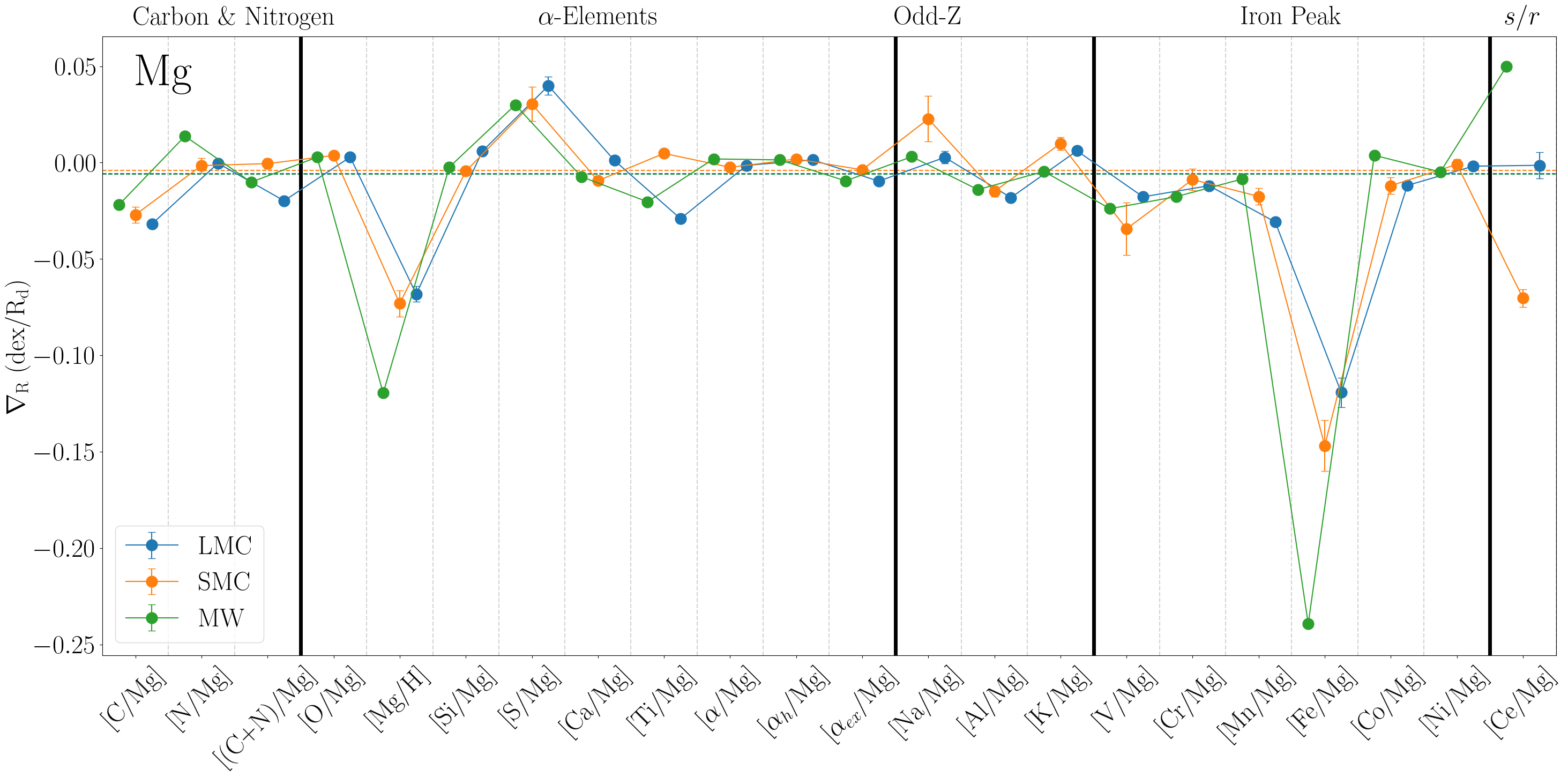}
    \caption{The same as Figure \ref{fig:smc_smc_lmc_mw}, but for the [X/Mg] gradients only. The median gradient value all three galaxies is quite similar.}
    \label{fig:smc_smc_lmc_mw_mg}
\end{figure*}



\bsp	
\label{lastpage}
\end{document}